\documentclass[aps,prd,twocolumn,showpacs,superscriptaddress,nofootinbib]{revtex4-2}

\usepackage{graphicx}
\usepackage{dcolumn}
\usepackage{bm}

\usepackage[english]{babel}
\usepackage{graphicx}
\usepackage{epsfig}
\usepackage{amssymb}
\usepackage{amsmath}
\usepackage[plainpages=false,pagebackref=false,pdftex]{hyperref}
\usepackage{placeins}
\hypersetup{colorlinks=true, linkcolor=blue, citecolor=blue,urlcolor=blue}
\usepackage{ulem}
\usepackage[T1]{fontenc} 
\usepackage[utf8]{inputenc}
\usepackage{color}
\usepackage[titletoc]{appendix}
\usepackage[percent]{overpic}

\begin{document}
\preprint{APS}

\title{Neutral pion masses within a hot and magnetized medium in a lattice-improved soft-wall AdS/QCD model}

\author{Nanxiang Wen }
\affiliation{ Department of Physics and Siyuan Laboratory, Jinan University, Guangzhou 510632, China}
\author{Xuanmin Cao}
\email[]{Corresponding author: caoxm@jnu.edu.cn}
\affiliation{ Department of Physics and Siyuan Laboratory, Jinan University, Guangzhou 510632, China}
\author{Jingyi Chao }
\email[]{Corresponding author: chaojingyi@jxnu.edu.cn}
\affiliation{ College of Physics and Communication Electronics, Jiangxi Normal University, Nanchang, Jiangxi 330022, China}
\author{Hui Liu}
\affiliation{ Department of Physics and Siyuan Laboratory, Jinan University, Guangzhou 510632, China}

\begin{abstract}
We  investigate chiral phase transitions and the screening masses, pole masses, and thermal widths of neutral pion meson with finite temperature $T$ and magnetic field $B$ in a lattice-improved AdS/QCD model, which is constructed by fitting the lattice results of the pseudo-critical temperatures \( T_{\text{pc}}(B) \). Specifically, we have that the chiral condensate \(\sigma\) undergoes a crossover phase transition demonstrating distinct magnetic catalysis and inverse magnetic catalysis effects in very low and high-temperature regions with fixed finite $B$, respectively. For the screening masses, we find that the longitudinal component decreases with $B$ at very low and high temperatures and increases with $B$ near $T_{\text{pc}}$. The transverse component always increases with $B$ at fixed $T$. However, both the longitudinal and transverse screening masses increase with $T$ at fixed $B$. Furthermore, we find that the pole mass decreases with the increasing of $B$ or $T$. Besides, it is interesting to note that the thermal width shows similar behavior to the longitudinal screening masses in the very high temperature region.
\end{abstract}

\maketitle
\tableofcontents

\section{Introduction}
The properties of strong interactions in the presence of a substantial external magnetic field have been studied extensively in recent years. The reason for this is that in off-center heavy-ion collisions, strong magnetic field of about \(10^{18}\) to \(10^{19}\) Gauss, or $|eB|\sim$ \(m_{\pi}^2\) to 15\(m_{\pi}^2\), are produced when two fast-moving nuclei collide~\cite{skokov2009estimate,deng2012event,voronyuk2011electromagnetic}. In the presence of such a strong external magnetic field, not only is the phase diagram of quantum chromodynamics (QCD) matter altered~\cite{chiralTcpIMClatticebali2012qcd,andersen2016phase,miransky2015quantum}, but also many novel and interesting phenomena will arise due to the interacting between the magnetic field and the non-perturbative properties of non-Abelian gauge theory, including the chiral magnetic effect~\cite{fukushima2008chiral,kharzeev2014chiral,kharzeev2008effects,kharzeev2011testing}, disputable superconductivity in magnetized vacuum~\cite{chernodub2010superconductivity,chernodub2011spontaneous}, neutral pion condensation~\cite{cao2016electromagnetic}, diamagnetism at low temperature and paramagnetism at high temperature~\cite{bali2014qcd}.

It has been shown that in a simplified effective theory, certain properties of the equilibrium plasma can be obtained by a reduction of the dimension from $D$ to $D-1$ at finite temperature. Furthermore, in the strong $B$ limit the longitudinal and transverse spaces are decoupled, allowing to reduce the transverse dimensions. Compared to magnetized matter at zero temperature, several unexpected phenomena have arisen, showing the interplay between the anisotropies induced by thermal excitation and the magnetic field. The most famous effect is that the chiral order parameter, characterized by spontaneous chiral symmetry breaking (SSB), is catalyzed by the magnetic field in the vacuum, known as magnetic catalysis (MC) in various models, including two-flavor QCD system~\cite{gusynin1996dimensional,buividovich2010numerical,braguta2012chiral}. However, the chiral pseudo-critical temperature $T_{\text{pc}}$ decreases with increasing $B$ in QCD matter. This is beyond the prediction of low-energy effective theories or model calculations, known as inverse magnetic catalysis (IMC)~\cite{chiralTcpIMClatticebali2012qcd,chiralMCandIMC1latticebali2012qcd}. In order to solve this puzzle, extensive work has been done by numerous approaches. (see recent reviews, e.g., Ref.~\cite{Andersen:2021lnk}). Among these explanations and calculations, the proper implementation of thermal modification is particularly important.

Therefore, we pay more attention to the properties of the light mesons at finite temperature. Since the pattern of SSB is $U(1)_{I_{3}}\otimes U(1)_{AI_{3}}\to U(1)_{L+R}$, the neutral pion is remained as the only Goldstone boson in two-flavor QCD. It plays a unique role in low-energy hadronic physics.~\cite{alkofer1989chiral,yulang1sheng2021pole,yulanglatticeNJLsheng2022impacts,latticepionding2022chiral,wen2023functional}. Our motivation for the study of thermal effects on $\pi^0$ is the observation that reliable information on the temperature modification of magnetized hadronic properties is still lacking. Since the temperature breaks the Lorentz invariance, it implies that one needs not only the pole mass $m_{\text{pole}}$, which describes the positions of the poles in the particle's propagators, but also the screening masses $m_{\text{scr}}$, which characterize the exponential decay of static propagators. The relationship between the screening mass and the pole mass is determined by the dispersion relation: $m_{\text{scr}}=m_{\text{pole}}/u$. In addition, there are two types of screening masses because the magnetic field breaks the $O(3)$ rotation symmetries to $O(2)$: \(m_{\text{scr},\parallel}\) and \(m_{\text{scr},\perp}\) for the masses along the direction of $B$ and those perpendicular to it, respectively. Required by the law of causality, as pointed out by ref.~\cite{yulang1sheng2021pole}, it can be deduced that \(u_{\perp,\parallel} < 1 \). Also, since motion along the transverse direction is suppressed compared to motion along the longitudinal direction, a naive estimate is that \(u_{\perp} < u_{\parallel} \). We conclude that \(u_{\perp} < u_{\parallel} < 1\) and so then \(m_{\text{scr},\perp} > m_{\text{scr},\parallel} > m_{\text{pole}}\)~\cite{yulang1sheng2021pole}. Over the decades, numerous models have been developed for studying the vacuum and thermal properties of hadrons with a magnetic field. These frameworks include Lattice QCD (LQCD)~\cite{ding2020meson,latticepionding2022chiral,Bali:2017ian,Bali:2017yku,Bignell:2020dze,Luschevskaya:2016epp},
chiral perturbation theory (\(\chi\)PT)~\cite{andersen2012chiral,Aung:2024qmf,Alvarado:2023loi}, Nambu--Jona-Lasinio models (NJL)~\cite{yulang1sheng2021pole,yulanglatticeNJLsheng2022impacts,pionmassdingxingwang2018meson,Fayazbakhsh:2013cha,Li:2020hlp},
functional renormalization group (FRG)~\cite{wen2023functional,kamikado2014chiral,Pawlowski:2023gym}. 

Besides the traditional methods, the discovery of the famous anti-de Sitter/conformal field theory (AdS/CFT) correspondence~\cite{4Dexternalsource1maldacena1999large,4Dexternalsource2gubser1998gauge,4Dexternalsource3witten1998anti} offers a powerful tool for investigating strongly coupling problems of QCD. In the framework of the bottom-up approach, several holographic QCD (HQCD) models has been constructed, e.g.   the hard-wall model~\cite{bottomup1erlich2005qcd} and soft-wall AdS/QCD model\cite{bottomup2karch2006linear} for chiral dynamics and hadronic physics, the Einstein-Maxwell-Dilaton system~\cite{gubser2008mimicking,gubser2008thermodynamics,gursoy2008exploring,gursoy2008exploring2} for thermodynamics, the light-front holographic QCD model~\cite{brodsky2015light} for hadronic physics, and so on. 

Among those HQCD models, we will take the soft-wall model as the start point of this work, since it can realize a good description on hadronic physics ~\cite{colangelo2008light,gherghetta2009chiral,kelley2011pseudoscalar,li2013dynamical,li2013dynamical2,sui2010prediction,ballon2020nonlinear,capossoli2020hadronic,Chen:2021wzj,Ahmed:2023pod,Guo:2023zjx} as well as the spontaneous chiral symmetry breaking \cite{laoshishouzhengNoB2-chelabi2016chiral,laoshishouzhengNoB1-chelabi2016realization}. Furthermore, the mass plane phase diagram from the soft-wall model~\cite{Li:2016smq,Chen:2018msc} are shown to be qualitatively consistent with the so-called ``Columbia plot''~\cite{Brown:1990ev,Ding:2015ona}. Since the model is constructed based on symmetry, it is quite direct to introduce the conserve current. Therefore, it is also convenient to extend the study to many different conditions, such as finite baryon/isospin densities \cite{Lv:2018wfq,Cao:2020ske,Chen:2019rez}, rotating mediums ~\cite{Chen:2022mhf}, none-quilibrium phase transitions ~\cite{Cao:2022mep}. Besides, the thermal properties as well as the Goldstone nature of pions has been systematically investigated in \cite{Cao:2022csq,laoshipionmassNoB1-cao2021thermal,xuanminprdjiezicao2020pion,Liang:2023lgs}.

Through introducing the magnetic field in the holographic models, it is possible to study the properties of QCD in the hot magnetized medium ~\cite{laoshishouzhengB1=li2017inverse,Rodrigues:2018pep,Rodrigues:2020ndy,fangzhenweakMCfang2016anomalous}. Actually, in ref.~\cite{laoshishouzhengB1=li2017inverse}, the authors obtained the IMC effect on chiral condensation and its pseudo-critical temperature $T_{\text{pc}}$. Additionally, result from ref.~\cite{fangzhenweakMCfang2016anomalous} show that chiral condensation not only exhibits IMC effect near $T_{\text{pc}}$ but also demonstrates MC effect at low temperatures. However, those studies are based on perturbative expansion solutions, and the magnetic field strength can not be extended to large values. Furthermore, those studies only analyze the order parameter, and the properties about the Goldstone bosons are absent. Therefore, in this work, we will try to solve the full magnetized background and study the thermal pions under this background. Furthermore, the variation of the chiral phase transition temperature $T_{\text{pc}}$ with $B$ in soft-wall AdS/QCD models differs from the lattice simulation results~\cite{chiralTcpIMClatticebali2012qcd,chiralMCandIMC1latticebali2012qcd}. Hence, we posit that those soft-wall AdS/QCD models should undergo alterations when introducing the magnetic field. Inspired by the the magnetic field-dependent four-fermion coupling constant $G(B)$ introduced in ref.~\cite{yulanglatticeNJLsheng2022impacts}, we will modify the 5D mass of the soft-wall AdS/QCD model by introducing a polynomial $g(B)$, which is considered as an effective coupling between the dilaton and the scalar field, or in some sense the coupling between the gluonic and chiral sector. We will fit such an effective coupling by comparing the data of $T_{\text{pc}}(B)$ with the lattice simulation~\cite{chiralTcpIMClatticebali2012qcd}. After fitting the model, we will systematically study the thermal properties of the neutral pions.

This paper is organized into the following sections. In Section~\ref{duguisec}, after introducing the Einstein-Maxwell system with back-reaction of magnetic field, we solve the Einstein-Maxwell equation and obtain the complete numerical background solutions with different temperatures $T$ and magnetic field strength $B$. In Section~\ref{shouzhengningjudaicichangxiaoying}, we modified the expression of the 5D mass $m_5$ in the AdS/QCD model through fitting the pseudo-critical temperatures $T_{\text{pc}(B)}$ from the Lattice simulation~\cite{chiralTcpIMClatticebali2012qcd}. Then, we numerically obtain the chiral condensates with different conditions of $T$ and $B$ and check out their magnetic catalysis and inverse magnetic catalysis effects. In Sections~\ref{scrandpole}, we extract and investigate the longitudinal screening masses $m_{\text{scr},\parallel}$, transverse screening masses $m_{\text{scr},\perp}$, pole masses $m_{\text{pole}}$, and thermal widths $\Gamma/2$ of the neutral pion under finite temperature and magnetic field from the lattice-improved AdS/QCD model. Finally, in Section~\ref{summaryanddiscussion}, we give a summary and discussion.

\section{Gravity background} \label{duguisec}
For the Gravity background, the authors in Ref.~\cite{laoshishouzhengB1=li2017inverse} employed an asymptotic expansion method to obtain an approximate solution of the Einstein-Maxwell (EM) system. However, the result is only reasonable under the condition $B \ll T^2$.~\footnote{In our study, the physical dimension of $B$ is GeV$^2$.} To extend to the large $B$ region, we will introduce the "shooting method" ~\cite{boyd2001chebyshev,laoshipionmassNoB1-cao2021thermal} to solve the EM system and obtain its full solution.

To introduce the magnetic field, we follow the strategy in refs.~\cite{yinruB1mamo2015inverse,yinruB2dudal2016no,laoshishouzhengB1=li2017inverse}. The back-reaction of the magnetic field is considered in the EM system. Its action in five-dimension space-time is 
\begin{equation}
   S_B=\frac{1}{16 \pi G_{5}}\int d^5x\sqrt{-g} \left(R-F_{MN} F^{{MN}}+\frac{12}{L^2} \right),
\end{equation}
with $R$ the scalar curvature, $L$ the AdS radius and $G_5$ the 5D Newton constant. The notations $M$, $N$ take values of $0, 1, \cdots, 4$. The $g$ is the metric determinant of $g_{MN}$. The $F_{MN}$ stands for a $U(1)$ gauge field. The matter part, described by the soft-wall AdS/QCD model, is considered as a probe on this background. Thus, the back-reaction from the matter
part is neglected.


\subsection{EOMs within the EM system}
Within the EM system, the equations of motion (EOMs) are
\begin{subequations}
\begin{equation}
   E_{MN}-\frac{6}{L^{2}} g_{MN} -\left(g^{IJ} F_{IM} F_{JN} -\frac{1}{4} F_{IJ} F^{IJ} g_{MN} \right)=0,
   \label{EMfcyuan}
   \end{equation}
   \begin{equation}
 \nabla _MF^{MN}=0
      \label{sanjiao},
\end{equation}
\end{subequations}
with $E_{MN}$ the Einstein tensor, $R_{MN}$ the $Ricci$ tensor and $R$  the $Ricci$ scalar, $E_{MN} = R_{MN} - \frac{R}{2}g_{MN}$. For satisfying the field equation for $F_{MN}$ in eq.~(\ref{sanjiao}), the configuration of a constant magnetic field could take as 
\begin{equation}
 F=\frac{B}{L} dx_1\land dx_2
      \label{cichangwaiji},
\end{equation}
where $L$ is set to unity in this work, $L=1$.

In this 5D coordinate system, it takes as $(t,x_1,x_2,x_3,z)$, in which the coordinate $z$ corresponds to the radial holographic direction. For the magnetic field align along the $x_3$ axis, one has $B_{x_3}=F_{x_1 x_2}=\partial_{x_1} A_{x_2}-\partial_{x_2}A_{x_1}=B$, where the bulk gauge potential is defined as $A_\mu(x_1,z)=\frac{1}{2}B(x_1\delta^{x_2}_{\mu}-x_2\delta^{x_1}_{\mu})$. As a result, the metric ansatz could be chosen as
\begin{equation} 
      ds^2=e^{2A(z)} \left(-f dt^2+\frac{1}{f}dz^2 +h \left(dx_{1}^{2}+dx_{2}^{2}\right)+q\,  dx_{3}^{2}\right), \label{duguijuzhen}
  \end{equation}
with $A(z)=-\text{ln}(z)$. As our study focuses on the thermal properties in the equilibrium state, the $f$, $h$, and $q$ all are functions only with respect to the variable $z$.

Under the given metric ansatz in eq.~\eqref{duguijuzhen}, EOM in eq.~\eqref{EMfcyuan} reduces to the following forms:
\begin{widetext}
\begin{subequations} \label{duguiquan}
\begin{equation}
    f''+f' \left(\frac{h'}{3 h}+\frac{q'}{6 q}-\frac{3}{z}\right)+f \left(-\frac{2 h' q'}{3 h q}+\frac{2 h'}{h z}-\frac{h'^2}{3 h^2}+\frac{q'}{q z}\right)-\frac{8 B^2 z^2}{3 h^2}=0,
\label{dugui1}
\end{equation}
\begin{equation}
    q''+q' \left(\frac{2 f'}{3 f}+\frac{h'}{3 h}-\frac{2}{z}\right)+q \left(-\frac{8 B^2 z^2}{3 f h^2}-\frac{2 f' h'}{3 f h}+\frac{2 h'}{h z}-\frac{h'^2}{3 h^2}\right)-\frac{q'^2}{2 q}=0,
\label{dugui2}
\end{equation}
\begin{equation}
    h''+h' \left(\frac{f'}{3 f}-\frac{q'}{6 q}-\frac{1}{z}\right)+h \left(\frac{q'}{q z}-\frac{f' q'}{3 f q}\right)-\frac{h'^2}{3 h}+\frac{4 B^2 z^2}{3fh}=0,
\label{dugui3}
\end{equation}
together with a constrain equation
\begin{equation} \label{dugui4}
   \frac{f' h'}{2 f h}+\frac{f' q'}{4 f q}-\frac{3 f'}{2 f z}-\frac{6}{f z^2}+\frac{h' q'}{2 h q}-\frac{3 h'}{h z}+\frac{h'^2}{4 h^2}-\frac{3 q'}{2 q z}+\frac{6}{z^2}+\frac{B^2 z^2}{f h^2}=0.
\end{equation}
\end{subequations}
\end{widetext}


\subsection{Complete numerical solutions}
\label{duguiquanjietu}
Through a careful analysis of the EOMs in eqs.~(\ref{duguiquan}), one can find that there are singularities at both $z=0$ and $z=z_h$, which result in the difficulties in getting analytical solutions. Nevertheless, we can employ the numerical algorithm "shooting method", as introduced in refs.~\cite{boyd2001chebyshev,laoshipionmassNoB1-cao2021thermal}, to solve these ODEs.

For eqs.~(\ref{duguiquan}), one can obtain the asymptotic expansions around the ultraviolet (UV) boundary at $z= 0$
\begin{subequations} \label{duguiz0}
\begin{eqnarray}
    f(z)&= 1+(\frac{2 B^2}{3 h_0^2})z^4 \ln(z)+f_4 z^4+\mathcal{O}(z^5),\;\;\;\;\; \\
    \label{zhankaifz0}
    q(z)&= q_0+(\frac{2q_0 B^2}{3h_0^2})z^4\ln{(z)}+q_4 z^4+\mathcal{O}(z^5),\,\\ \label{zhankaiqz0}
    h(z)&= h_0-(\frac{B^2}{3h_0})z^4\ln{(z)}-\frac{h_0 q_4}{2q_0}z^4+\mathcal{O}(z^5) ,
    \label{zhankaihz0}
\end{eqnarray}
\end{subequations}
with $f_4$, $q_0$, $q_4$ and $h_0$ the integration constants.

At the horizon $z=z_{h}$ or the infrared (IR) boundary, in order to get black hole solution, the condition $f(z=z_{h})=0$ must be satisfied. Thus, for the asymptotic expansion of $f(z)$ at $z= z_h$, there is no constant term. Near the horizon, one can derive the asymptotic expansions of eqs.~(\ref{duguiquan}) as
\begin{subequations} \label{duguizh}
\begin{eqnarray}
f(z)=&&f_{h1}(z-z_h)+\frac{1}{3 z_h^2}\Big(\frac{5 B^2 {z_h}^4}{h_{{h0}}^2}+3 {z_h} f_{{h1}}-6 \Big)\nonumber\\
&&\times(z-z_h)^2+\mathcal{O}(z-z_h)^3,  \\  
\label{zhankaifzh}
q(z) =&& q_{h0}+\frac{2 q_{{h0}} }{3 f_{{h1}} h_{{h0}}^2 z_h^2}\nonumber\\
&&\times\left(2 B^2 z_h^4+3 f_{{h1}} h_{{h0}}^2 z_h+12 h_{{h0}}^2\right)(z-z_{h})\nonumber\\
&&+\frac{q_{h0}}{9 {z_h}^4 f_{{h1}}^2 h_{{h0}}^4} \Big[20 B^4 {z_h}^8+12 B^2 {z_h}^5 f_{{h1}} h_{{h0}}^2\nonumber\\
&&+9 h_{{h0}}^4 \left({z_h} f_{{h1}}+4\right){}^2 \Big](z-z_h)^2+\mathcal{O}(z-z_h)^3 ,\nonumber\\
\\
\label{zhankaiqzh}
h(z) =&& h_{h0}+\frac{2}{3 f_{{h1}} h_{{h0}} z_h^2}\nonumber\\
&&\times\left(-4 B^2 z_h^4+3 f_{{h1}} h_{{h0}}^2 z_h+12 h_{{h0}}^2\right)(z-z_{h})\nonumber\\
&&+\frac{1}{9 {z_h}^4 f_{{h1}}^2 h_{{h0}}^3} \Big[8 B^4 {z_h}^8-24 B^2 {z_h}^4 h_{{h0}}^2 \nonumber\\
&&\times\left({z_h} f_{{h1}}+3\right)+9 h_{{h0}}^4 \left({z_h} f_{{h1}}+4\right){}^2 \Big]\nonumber\\
&&\times(z-z_h)^2+\mathcal{O}(z-z_h)^3 ,
    \label{zhankaihzh}
\end{eqnarray}
\end{subequations}
where the $f_{h1}$, $q_{h0}$ and $h_{h0}$ are the integration constants.

\begin{figure*}[htb]
\centering
    \begin{overpic}[scale=0.4]{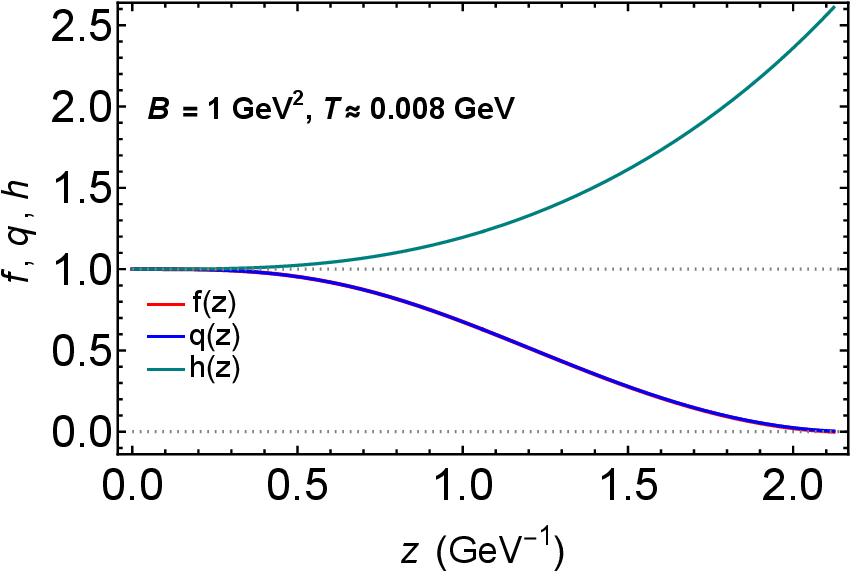}
        \put(25,60){\bf{(a)}}
    \end{overpic}
    \begin{overpic}[scale=0.4]{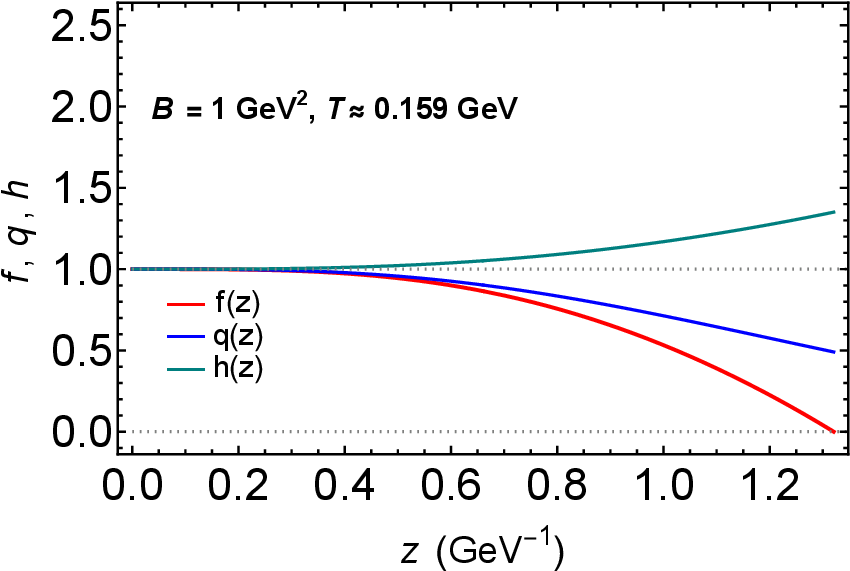}
        \put(25,60){\bf{(b)}}
    \end{overpic}
     \begin{overpic}[scale=0.4]{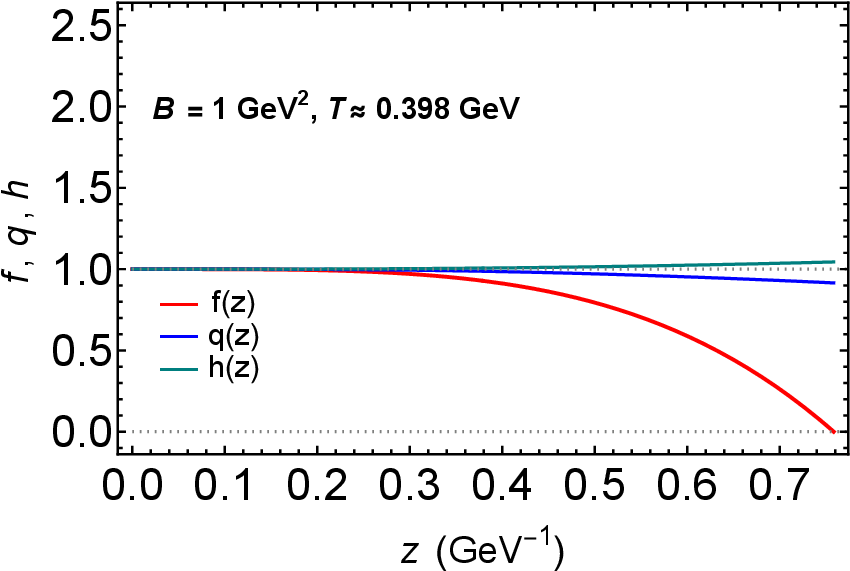}
        \put(25,60){\bf{(c)}}
    \end{overpic}
     \begin{overpic}[scale=0.4]{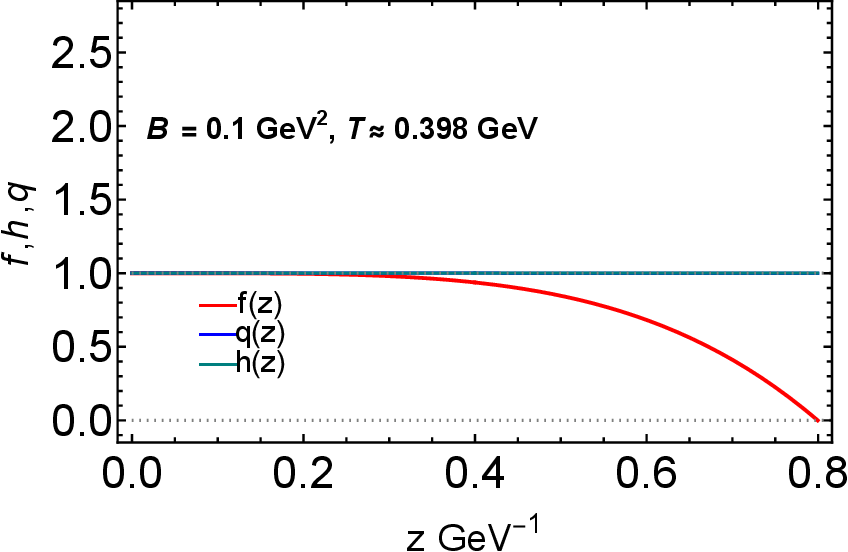}
        \put(25,60){\bf{(d)}}
    \end{overpic}
    \begin{overpic}[scale=0.4]{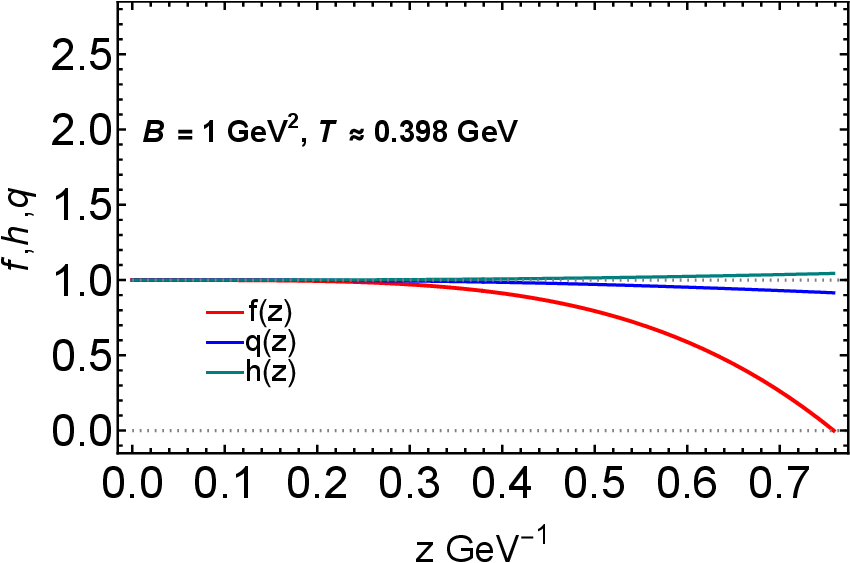}
        \put(25,60){\bf{(e)}}
    \end{overpic}
     \begin{overpic}[scale=0.4]{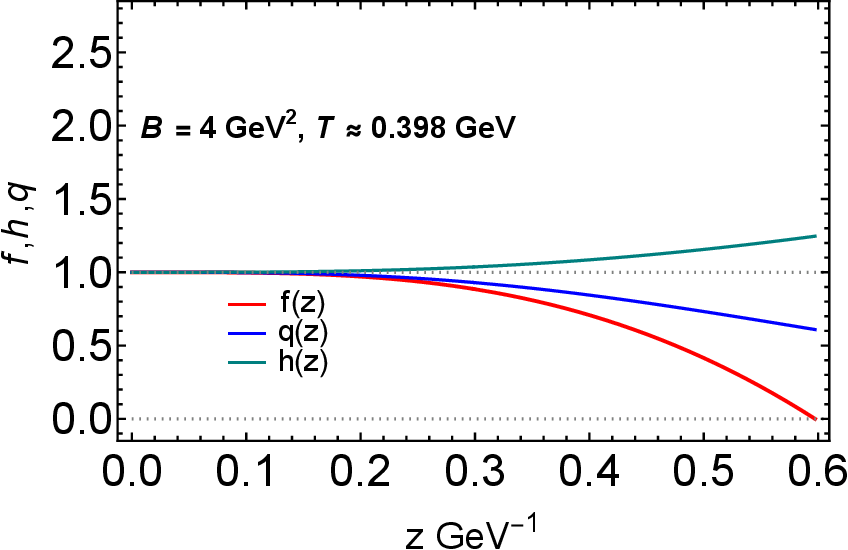}
        \put(25,60){\bf{(f)}}
    \end{overpic}
    \begin{overpic}[scale=0.41]{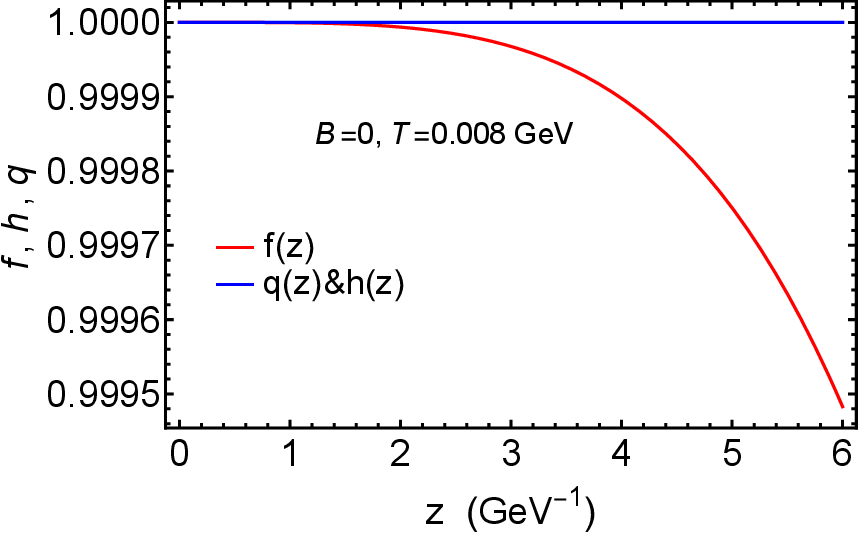}
        \put(25,50){\bf{(g)}}
    \end{overpic}
    \begin{overpic}[scale=0.4]{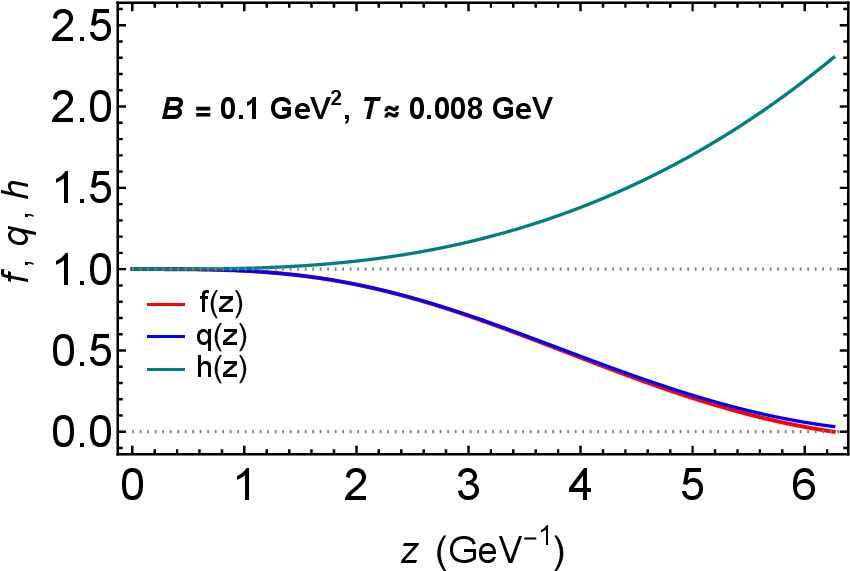}
        \put(25,60){\bf{(h)}}
    \end{overpic}
     \begin{overpic}[scale=0.4]{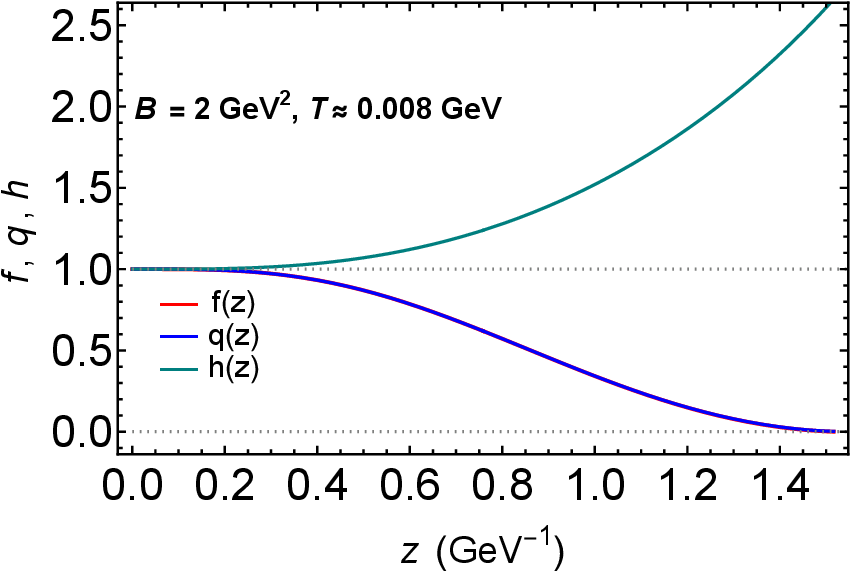}
        \put(25,60){\bf{(i)}}
    \end{overpic}
\caption{The $f$, $q$, and $h$ as functions of $z$ are represented by the lines of red, blue, and cyan, respectively. (a)-(c), magnetic field $B$ is fixed at 1 GeV$^2$. Temperature $T$ is fixed at 8, 159, 398 MeV, respectively. (d)-(f), temperature $T$ is fixed at high value, $T\approx398$MeV. Magnetic field $B$ is fixed at 0.1, 1, 4 GeV$^2$, respectively. (h)-(i), temperature $T$ is fixed at low value, $T\approx 8$MeV. Magnetic field $B$ is fixed at 0, 0.1, 2 GeV$^2$, respectively.}
\label{fqhwithzonTTTandBBB}
\end{figure*}

Physically, when the magnetic field is absent, $B=0$, the metric in eq.~(\ref{duguijuzhen}) should reduce into
\begin{equation} 
      dS^2=\frac{L^2}{z^2} \left(-f dt^2+\frac{1}{f}dz^2 + dx_{1}^{2}+dx_{2}^{2}+ dx_{3}^{2}\right).
      \label{duguijuzhen0}
  \end{equation}
This change stems from the restoration of isotropy within the spatial dimensions $x_1, x_2$ and $x_3$. From eq.~(\ref{duguijuzhen0}), it becomes evident that $h=q=1$ at $B=0$. By Considering the zero magnetic field limit in the asymptotic expansions eq.~\eqref{duguiz0}, it is obviously integral constants \(q_0\) and \(h_0\) should satisfy $q_0=1$ and $h_0=1$. Furthermore, the temperature is related to the horizon
\begin{equation}
    T=\Big|\frac{f'(z)}{4\pi}\Big|_{z=z_h}.
    \label{wenduzhguanxi}
\end{equation}
Thus, from eqs.~\eqref{zhankaifzh} and \eqref{wenduzhguanxi}, one can find that $z_h$ is a function of $B$ and $T$ (or the temperature-related quantity $f_{h1}$).

Within this system, considering the eq.~(\ref{dugui4}), the number of constraints is reduced to five.
However, considering the asymptotic expansions of eqs.~(\ref{duguiquan}) at $z=0$ and $z=z_h$, the total count of involved unknown integral constants amounts to 5, which are $f_4$ and $q_4$ at $z=0$ boundary, $f_{h1}$, $q_{h0}$ and $h_{h0}$ at horizon.

In mathematics, within this EM system, the integral constants of the EM system and $z_h$ should be clearly determined by the given values of \(B\) and \(f_{h1}\). Thus, in our solving process, we treat the $f_4$, $q_4$, $q_{h0}$, $h_{h0}$ and $z_h$ as unknown quantities for any given $B$ and temperature-related quantity $f_{h1}$.

\begin{figure*}[htb]
\centering
\begin{overpic}[scale=1]{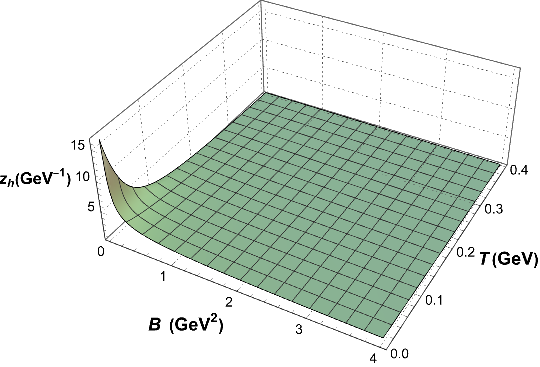}
        \put(20,1){\bf{(a)}}
    \end{overpic}~~~~~~~~~~~~~
    \begin{overpic}[scale=0.7]{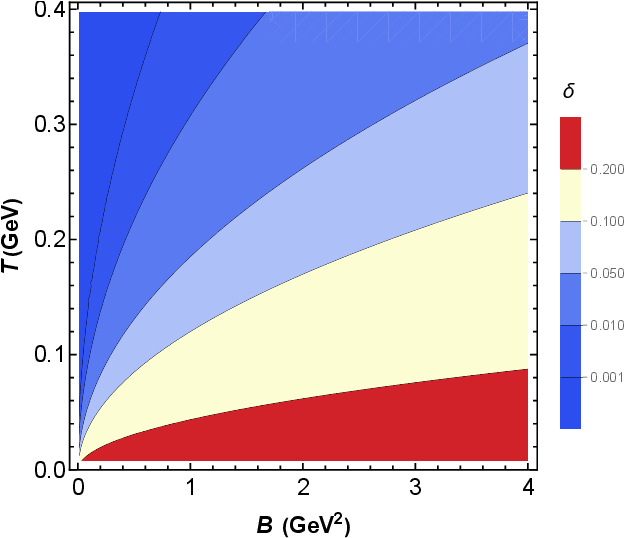}
        \put(20,1){\bf{(b)}}
    \end{overpic}
\caption{(a), the horizon boundary \(z_h\) as a function of temperature $T$ and magnetic field $B$. (b), the distribution of relative error of $z_h$, $\delta=\Delta/z_h=(z_h-\tilde{z}_h)/z_h$ with $\tilde{z}_h$ the perturbation result, between the results from the perturbative asymptotic approximation and the full solution across the plane of the temperature $T$ and magnetic field $B$.}
\label{duguiwanquantu}
\end{figure*}

To present how the temperature and magnetic field affect the background metric, we show variation curves of $f(z)$, $q(z)$ and $h(z)$ in Fig.~\ref{fqhwithzonTTTandBBB}. In Fig.~\ref{fqhwithzonTTTandBBB}(a-c), without loss of generality, we choose the $B$ is fixed at 1 GeV$^2$ as an example to check the temperature effects with $T\approx 0.008$, $0.159$ and $0.398$ GeV$^2$ in (a), (b) and (c), respectively. We can find that, as the temperature increases, the difference between $q(z)$ and $h(z)$ is vanishing, and both of them tend to unity. Considering the zero magnetic field limit metric in eq.~(\ref{duguijuzhen}), this implies that an increase in temperature is beneficial for restoring the spatial rotation invariance broken by the magnetic field.~\footnote{In fact, in our calculations, no matter what value the $B$ is fixed at, the temperature will always promote spatial isotropy. In physically, in higher temperature regions, the whole system will appear more chaotic. High temperatures lead to a large number of inelastic collisions, causing energy and momentum transfer between various degrees of freedom. Therefore, high temperatures are beneficial to "smoothen" the anisotropy brought by the magnetic field.} 
In Fig.~\ref{fqhwithzonTTTandBBB}(d-f), we show the numerical results at fixed temperature $T\approx 0.398$ GeV to present the affect of the variation of magnetic field. we observe that even at very high temperature $T\approx$ 0.398 GeV, the curves of $q(z)$ and $h(z)$, which is almost collapsed at small $B$,  become splitting  as the increasing of $B$. This indicates that  the increasing of the magnetic field enhances the spatial anisotropy. Therefore, there exist competitive between the temperature and magnetic field for the spatial anisotropy of the background. On the one hand, increasing the magnetic field enhances the spatial anisotropy. On the other hand, increased temperature favors the restoration of spatial asymmetry.

Moreover, as shown in Fig.~\ref{fqhwithzonTTTandBBB}(g-i), at very low fixed temperature $T \approx 8$ MeV, as the increasing of $B$, $q(z)$ and $h(z)$ will split, while $f(z)$ and $q(z)$ will converge. This suggests that at very low temperatures and large magnetic field, only slight difference exists between the temporal direction and $x_3$-direction. According to the trend, it might be that it will be identical in the $x_3$-direction and the temporal one at $T=0$. The physic in $x_3$-direction and temporal one satisfy Lorentz rotation invariance. Moreover, in ref.~\cite{yulang1sheng2021pole}, it argued that when $T=0$, $B>0$, there exist $m_{\text{pole}} = m_{\text{scr},\parallel}$. This is consistent with our deduction.

In Fig.~\ref{duguiwanquantu}(a), we show the $T$ and $B$ dependence of $z_h$ where the magnetic field $B$ is from 0.01 $\text{GeV}^2$ to 4 $\text{GeV}^2$ and the temperature $T$ is from 0.008 $\text{GeV}$ to 0.4 $\text{GeV}$. Notably, at lower $T$ and $B$, the value of $z_h$ is significant large. With increasing $T$ and $B$, the $z_h$ undergoes a sharp decrease. Furthermore, we have a compare to the perturbative asymptotic approximation results given in ref.~\cite{laoshishouzhengB1=li2017inverse}. In Fig.~\ref{duguiwanquantu}(b), we present the relative error (RE) of \(z_h\) between full solutions and the perturbative asymptotic one. The comparison shows that the perturbative asymptotic approximation agree well with the full solution outcomes in the region where $B \ll T^2$. Nevertheless, the deviations especially emerge at that region of low values of $T$ and high values of $B$. Hence, when we do research at the low-temperature and high-magnetic-field regime, the full solution is necessary to take and the asymptotic approximation is not suitable.

\section{The IMC and MC effects on chiral phase transition} \label{shouzhengningjudaicichangxiaoying}
After obtained the full solution for the EM system, we can explore the chiral phase transition through the order parameter (the chiral condensation).
Among the various models investigating chiral phase transitions, the soft-wall model stands out as one of the most effective models. And we investigate the chiral properties based on the soft-wall model.

In ref.~\cite{laoshishouzhengB1=li2017inverse}, chiral condensation exhibits noticeable IMC behavior and its $T_{\text{pc}}$ decrease with increasing $eB$. Under this model, we numerically calculate the full background solution and the chiral condensation. The results indicate that no MC effect of chiral condensation exists at low temperatures. However, the LQCD simulations~\cite{chiralTcpIMClatticebali2012qcd,chiralMCandIMC1latticebali2012qcd,latticepionding2022chiral} suggest that the chiral condensation should exhibit distinct MC and IMC behaviors at low temperatures and near $T_{\text{pc}}$ respectively.  The $T_{\text{pc}}(B)$ monotonously decreases with the increasing of $B$. Additionally, in IR-improved soft-wall AdS/QCD model, the results indicate that the MC effects of chiral condensation at low temperatures are very weak~\cite{fangzhenweakMCfang2016anomalous}. Therefore, to more consistently describe the MC and IMC effects, it might be necessary to incorporate the magnetic field effects within the soft-wall model. Inspired by ref.~\cite{yulanglatticeNJLsheng2022impacts}, we modify the 5D mass of the soft-wall AdS/QCD model by introducing a polynomial $g(B)$. The $g(B)$ is constrained by the variation of the $T_{\text{pc}}$ with $B$ from lattice simulations~\cite{chiralTcpIMClatticebali2012qcd}. In ref.~\cite{Liang:2023lgs}, they attempt to consider coupling between the dilaton $\Phi$ and scalar field $X$. Furthermore, in this work, we further optimized the AdS/QCD model through introducing a magnetic field-dependent polynomial $g(B)$ into a effective 5D mass expression.

The soft-wall AdS/QCD model is constructed within the bottom-up framework \cite{bottomup1erlich2005qcd, bottomup2karch2006linear} through considering the $SU(N_f)_L\times SU(N_f)_R$ gauge symmetry.  The action is expressed as
\begin{eqnarray}
   S_M=&& \int dz \int d^4x \sqrt{g} e^{-\Phi(z)} \text{Tr} \{|DX|^2-m_{5}^2(z)|X|^2\nonumber\\
   &&-\lambda|X|^4-\frac{1}{4 g_5^2}(F_L^2+ F_R^2)\}, 
   \label{zuoyongliang}
\end{eqnarray}
with $g$ the determinant of the metric $g_{MN}$. And $g_5=2\pi$, which is determined by comparing the large momentum expansion of the correlator of the vector current $J^a_\mu=\overline{q} \gamma_\mu t^a q$ in both the AdS/QCD and perturbative QCD~\cite{bottomup1erlich2005qcd}. The dilaton profile, $\Phi(z)=\mu_g^2z^2$, with $\mu_g$ the constant that is necessary for the Regge behavior of the meson spectrum~\cite{bottomup2karch2006linear}. The covariant derivative $D^M$ and the field strength $F^{MN}_{L,R}$ are defined as $D^MX=\partial^MX-i A_{L}^{M}X+iXA_R^M$ and $F^{MN}_{L,R}=\partial^M A^N_{L,R}-\partial^N A^M_{L,R}-i[A^M_{L,R},A^N_{L,R}]$  with $A^M_{L,R}=A^{a,M}_{L,R}t^a_{L,R}$. The $t^a_{L,R} (a=1,2,3)$ is defined as the generators of $SU(2)_L$ and $SU(2)_R$ respectively with $M,N=0,1,2,3,4$. In this work, we only consider the case of $N_f=2$ with equal $u$ and $d$ quark mass, $m_q=m_u=m_d$.

For the sake of convenience, the gauge field $A_{L/R}$ can generally be rearranged into the vector field $V^M=\frac{1}{2}(A^M_L+A^M_R)$, and the axial-vector field $A^M=\frac{1}{2}(A^M_L-A^M_R)$. The vector field $V^M$ correspond as dual entities to the vector current $J_\mu^V$. The axial-vector field $A^M$ correspond as dual entities to the axial-vector current $J_\mu^A$.  Accompanied by the transformed chiral gauge field strengths denote as
\begin{subequations}
    \begin{eqnarray}
        F^{MN}_A&&=\frac{1}{2}(F^{MN}_L-F^{MN}_R) \\
        &&= \partial^MA^N-\partial^NA^M-i[V^M,A^N]-i[A^M,V^N] \nonumber, 
    \end{eqnarray}
      \begin{eqnarray}
        F^{MN}_V&&=\frac{1}{2}(F^{MN}_L+F^{MN}_R) \\
        &&= \partial^MV^N-\partial^NV^M-i[V^M,A^N]-i[A^M,V^N] \nonumber. 
    \end{eqnarray}
Furthermore, the covariant derivative is
\begin{eqnarray}
    D_MX=\partial_MX-i[V_M,X]-i\{A_M,X\}.
\end{eqnarray}
\end{subequations}

Considering the coupling between the scalar field $X$ and dilaton $\Phi$ $m_5^2(z)$ , it effectively lead to the modification of $m_5$,
\begin{equation} \label{orange5dd}
    m_5^2(z)=-3 \big[1+\gamma  \tanh (\kappa  \Phi) \big],
\end{equation}
with $\gamma, \kappa$ the free parameters. 
The leading constant term for the 5D mass, denoted as $m_5^2=-3$, can be established through the AdS/CFT dictionary: $m_5^2 = (\Delta - p)(\Delta + p - 4)$, with the choice of $p = 0$ and $\Delta = 3$, where $\Delta$ represents the dimension of the dual operator $\overline{q}_R \overline{q}_L$ \cite{bottomup1erlich2005qcd}.

In ref.~\cite{Liang:2023lgs}, the chiral condensation has a slight unphysical bumps up at low temperatures region with the increasing temperature, so that the model parameters should be adjusted. Our goal is to achieve the pion mass $M_{\pi}$~\footnote{The uppercase \(M_{\pi}\) in this article specifically denotes the mass of the \(\pi\) meson under conditions of zero temperature, zero density, and zero magnetic field strength.}, \(M_{\pi}\approx 139.6 (\text{MeV})\)~\cite{ParticleDataGroup:2014cgo}, and the saturation chiral condensate $\sigma$, $\sigma\approx 0.0278 (\text{GeV}^3)$ ~\cite{li2013dynamical} . Details of the new parameters are provided in Table.~\ref{canshuNoeB}. The selection and adjustment of parameters involve following steps. Initially, we adjusted the value of \(\mu_g\), a parameter associated with the Regge behavior of the meson spectrum, and fixed it at 0.35 GeV. Subsequently, we adjusted the magnitude of \(\lambda\), which directly influences the value of the chiral condensation \(\sigma\), setting it to 25. Finally, we determined the values of \(\kappa\) and \(\gamma\). In our parameter tuning, we found that the values of \(\kappa\) and \(\gamma\) not only affect the $T_{\text{pc}}$ but also impact the magnitude of \(\sigma\). In the absence of an external magnetic field, given that the pseudo-critical temperature \(T_{\text{pc}}\) should lie between 150 and 170 MeV, we fit \(\kappa=0.85\) and \(\gamma=6\). Additionally, as the quark mass \(m_q\) is positively correlated with \(M_{\pi}\), we fit \(m_q\) to 3.9 MeV. The computed result with these parameters yields the pseudo-critical temperature \(T_{\text{pc}}(B=0)\approx 161\) MeV.

Furthermore, during the parameter adjustment process, we observed that changing the coefficient of the dilation function \(\Phi(z)\) in eq.~(\ref{orange5dd}) alone could affect both the values of $\sigma$ and the \(T_{\text{pc}}\). Therefore, when introducing the magnetic field-dependent function \(g(B)\) into the effective 5D mass in eq~(\ref{orange5dd}), we have $\kappa+g(B)$ as the coefficient of the dilation \(\Phi(z)\). Finally, to make the coefficient dimensionless, we have a reduced $B$ as \(B/\mu_g^2\). After incorporating the magnetic field-dependent polynomial $g(B)$, the expression for the effective 5D mass becomes
\begin{eqnarray}
     m_5^2(z)=-3 \bigg\{1+\gamma  \tanh \Big[\Big(\kappa+g(B)\Big)  \Phi \Big] \bigg\},
\end{eqnarray}
where
\begin{eqnarray}
    g(B)=\sum_i \eta_i\bigg(\frac{B}{\mu_g^2}\bigg)^{\alpha_i},
\end{eqnarray}
with $\alpha_i=2,4,6,8,10$. Since the magnitude of chiral condensation should not be impacted by the direction of the magnetic field, we set all \(\alpha_i\) to be even. In order to ensure that the variation of the \(T_{\text{pc}}(B)\) falls entirely within the error band of LQCD result, we adjusted \(g(B)\) based on the \(T_{\text{pc}}\) data points extracted from LQCD simulations~\cite{chiralTcpIMClatticebali2012qcd}. Furthermore, we observed that it is different in \(M_{\pi}\) and \(T_{\text{pc}}(B=0)\) for different effective models. Thus, we opted to use a dimensionless quantity \(B/M_{\pi}^2\) to measure the strength of the magnetic field. We also used the normalized pseudo-critical temperature \(T_{\text{pc}}(B)/T_{\text{pc}}(B=0)\) to assess the variation in the pseudo-critical temperature with $B$. Finally, we get the fitted coefficients as shown in Table~\ref{canshugeB}.

\begin{table}[b]
\caption{\label{canshuNoeB}
Parameters of Model.}
\begin{ruledtabular}
\begin{tabular}{cccccc}
Parameters & $m_q(\text{GeV})$ & $\mu_g(\text{GeV})$ & $\kappa$ & $\gamma$ & $\lambda$ \\
\hline
Value & $3.90\times10^{-3}$ & $0.35$ & $0.85$ & $6$ & $25$
\end{tabular}
\end{ruledtabular}

\caption{\label{canshugeB}
Coefficients of $\eta_i$.}
\begin{ruledtabular}
\begin{tabular}{cccccc}
& $\eta_1$ & $\eta_2$ & $\eta_3$ & $\eta_4$ & $\eta_5$ \\
\hline
& $0.021$ & $-8.65 \times 10^{-4}$ & $2.02 \times 10^{-5}$ & $-2.37\times 10^{-7}$ & $1.10\times 10^{-9}$
\end{tabular}
\end{ruledtabular}
\end{table}

\begin{figure*}[htb]
\centering
    \begin{overpic}[width=0.48\textwidth,height=0.3\textwidth]{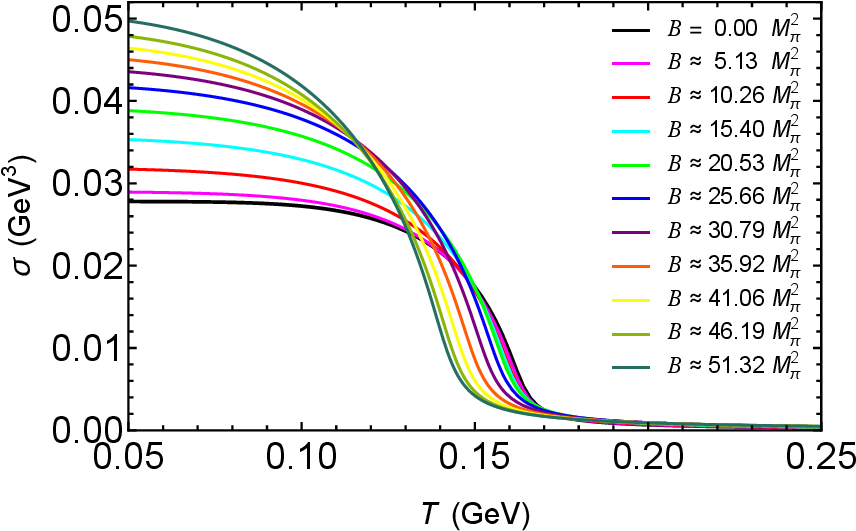}
        \put(45,58){\bf{(a)}}
    \end{overpic}
    \begin{overpic}[width=0.48\textwidth,height=0.3\textwidth]{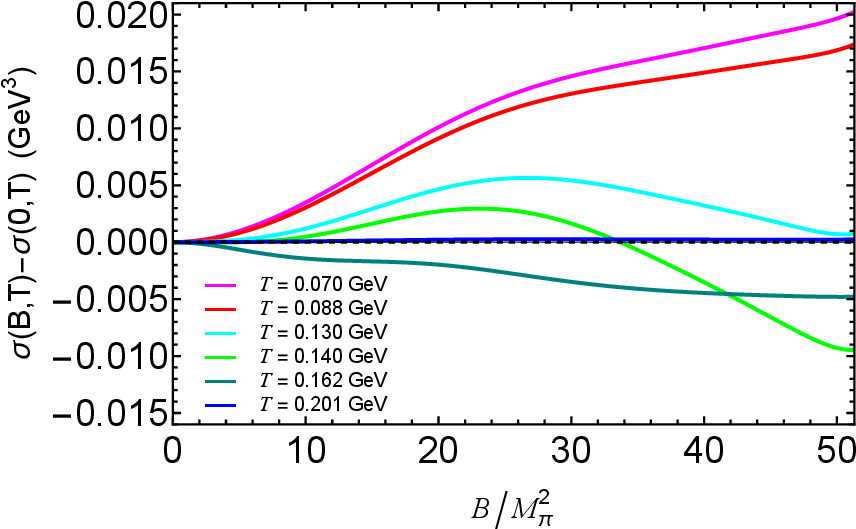}
        \put(45,58){\bf{(b)}}
    \end{overpic}
\caption{(a), the plot illustrates the chiral condensation \(\sigma\) as a function of \(T\) under different fixed \(B\). The range of \(B\) is approximately from 0 to 51.32 \(M_{\pi}^2\), which corresponds to 0 to 1 GeV\(^2\) in our model. (b), it depicts \(\Delta \sigma=\sigma(B,T)-\sigma(0,T)\) as a function of \(B/M_{\pi}^2\) under different fixed \(T\), where the temperatures \(T\) were chosen as 70, 88, 130, 140, 162, and 201 MeV, respectively.}
\label{ciralcondensate}
\end{figure*}

\begin{figure}[htb]
\centering
    \begin{overpic}[scale=0.55]{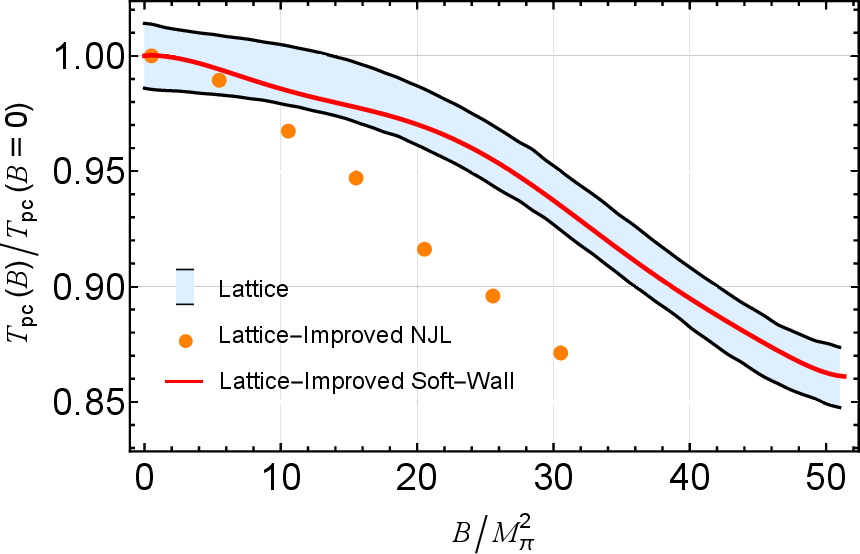}
    \end{overpic}
\caption{The pseudo-critical temperature scaled by its $B=0$ value as a function of the $B/M_{\pi}^2$. These results compare with those from lattice simulations~\cite{chiralTcpIMClatticebali2012qcd} and lattice-improved NJL~\cite{yulanglatticeNJLsheng2022impacts}. Given that the value of \(M_{\pi}\) in lattice simulations and the lattice-improved NJL model are 135 and 138 MeV, respectively, the results have been scaled accordingly for their respective \(M_{\pi}\) in the extracted data.}
\label{NormalizedTcpof2model}
\end{figure}


In the $N_f=2$ case, with equal $u$ and $d$ quark mass, $m_q=m_u=m_d$, the bulk scalar field X can be expressed as
\begin{equation}\label{expressofX}
X=(\frac{\chi}{2}+S)I_2e^{2i\pi^a t^a},
\end{equation}
where $\chi(z)$ is linked to the vacuum expectation value (VEV) of $X$ and $I_2$ represents the $2\times2$ identity matrix. Therefore,
one can derive the EOM for $\chi$ as
\begin{eqnarray} \label{chieom}
    \chi''+\bigg(3A'-\Phi'+ && \frac{f'}{f}+\frac{h'}{h}+\frac{q'}{2q} \bigg)\chi' \nonumber \\
    && - \frac{e^{2A}\chi \left(\lambda \chi^2+2m_5^2 \right)}{2f}=0.
\end{eqnarray}

The EOM for $\chi$ in eq.~\eqref{chieom} is a nonlinear second order differential equation with singularities both at $z=0$ and $z=z_h$, which makes analytical solutions difficult to attain. Nevertheless, we can numerically solve it with a algorithm (``Shooting mehtod'') as utilized in ref.~\cite{laoshipionmassNoB1-cao2021thermal}. Within this algorithm, by considering the eqs.~(\ref{duguiquan}), we can derive the asymptotic expansions of the $\chi$ at both the UV and IR boundaries as
\begin{subequations} \label{phim5chi}
    \begin{eqnarray}
        \chi(z\rightarrow 0)=&& m_q \zeta z+\frac{\sigma}{\zeta}z+\frac{1}{4}  m_q\zeta \Big[-6 \gamma  \mu _g^2 g(B)\\
        &&+(4-6 \gamma  \kappa ) \mu _g^2+\lambda  \zeta ^2 m_q^2 \Big] z^3 \ln(z)+\mathcal{O}(z^4) \nonumber,
    \end{eqnarray}
    \begin{eqnarray}
        \chi(z\rightarrow z_h)=&& \chi_{zh0}+\frac{1}{2 f_{h1} z_h^2}\chi_{zh0}\Big\{\lambda  \chi _{zh0}^2-6-6 \gamma \nonumber \\
        &&\times\tanh \Big[\big(\kappa + g(B)\big)\mu _g^2 z_h^2\Big]\Big\}(z-z_h)\nonumber\\
        &&+\mathcal{O}(z-z_h)^2,
    \end{eqnarray}
\end{subequations}
with integration constants $m_q$, $\sigma$ and $\chi_{zh0}$.
According to the holographic dictionary, the two integral constants, $m_q$ and $\sigma$, correspond to the quark mass and the chiral condensate $\sigma \equiv \left \langle \overline{q}q \right \rangle$, respectively. Furthermore, the normalization constant $\zeta$ takes the value $\sqrt{N_c}/2\pi$ with $N_c=3$, which is determined by matching to 4D QCD~\cite{quanxisuanziqq-cherman2009chiral}.

In Fig.~\ref{ciralcondensate}, we show the variation of chiral condensation with \(T\) (or \(B\)) when \(B\) (or \(T\)) is fixed. In Fig.~~\ref{ciralcondensate}(a), under different fixed \(B\) and varying $T$, the chiral phase transition exhibits crossover behavior. The results suggest that chiral condensation displays evident MC effects at low temperatures and significant IMC effects near \(T_{\text{pc}}(B)\). From Fig.~\ref{ciralcondensate}(b), on one hand, we observe that chiral condensation exhibits MC effects at low temperature. As the temperature increases to 140 MeV, chiral condensation initially shows MC behavior. Then, when the magnetic field strengths \(B\) is at $B\gtrsim$ 34 \(M_{\pi}^2\), the behavior of chiral condensation transitions shift from MC to IMC effect. Finally, when the temperature reaches 162 MeV, chiral condensation completely manifests IMC effects. These findings are qualitatively consistent with lattice simulation results~\cite{chiralTcpIMClatticebali2012qcd,chiralMCandIMC1latticebali2012qcd}. On the other hand, when the temperature is at $T\gtrsim$ 201 MeV, the variation of \(\Delta \sigma\) is almost decoupled from the magnetic field.

As shown in Fig.~\ref{NormalizedTcpof2model}, we have also derived the results for the normalized pseudo-critical temperature $T_{\text{pc}}(B)/T_{\text{pc}}(0)$. The pseudo-critical temperature is defined as $(\partial^2\sigma/\partial T^2)|_{T=T_{\text{pc}}}=0$. We can observe that the normalized \(T_{\text{pc}}(B)\) results fall entirely within the error band of lattice simulations~\cite{chiralTcpIMClatticebali2012qcd}. Additionally, we extracted results of $T_{\text{pc}}$ from the lattice-improved NJL model~\cite{yulanglatticeNJLsheng2022impacts} for comparing.

\section{Correlation functions and the masses of neutral pion at finite temperature and magnetic field} \label{scrandpole}
In the previous section, we construct the lattice-improved soft-wall AdS/QCD model. We use this model to study the chiral condensate and the chiral phase transition. In this section, we focus on the calculation of the screening masses~$m_{\text{scr}}$, the pole masses~$m_{\text{pole}}$ and the thermal widths~$\Gamma$ of the neutral pion $\pi^0$ under different magnetic field strengths $B$ and temperatures $T$.

The screening mass $m_{\text{scr}}$ describe the exponential decay of the spatial correlator $G(\vec{x})\sim e^{-m_{\text{scr}} |\vec{x}|}/|\vec{x}|$. Or it is the pole in the momentum space while the correlation function has the following form near the pole
\begin{equation} \label{screenyinyong}
    G(0,\textit{\textbf{k}})\sim \frac{1}{\textit{\textbf{k}}^2+m_{\text{scr}}^2}\,.
\end{equation}
Besides the spatial components, the information of the temporal component is also important for getting a full understanding on the mesonic correlation. It is described by the pole mass and the thermal width. Both of them  appear in the pole in complex frequency plane, near which the temporal correlation function takes the form 
\begin{equation} \label{poleyinyong}
    G(k_{0},\textbf{0})\sim \frac{1}{k_0-(m_{\text{pole}}-i\Gamma/2)}\,.
\end{equation}
Here we take the four vector $k$ as $k=(k_0,\textit{\textbf{k}})$, and $k_0,\textit{\textbf{k}}$ represent the frequency and the spatial momentum respectively.

Thus, to get those quantities, one has to calculate the correlators. Through the holographic approach, a linkage is forged between the 4D operator $\hat{O}(x)$ and the 5D field $\phi_0(x,z)$ by equating their partition functions. This methodology stands as a robust means to tackle strong coupling correlation functions
\begin{equation} \label{5D4Dlianxi}
    \left<e^{i \int d^4x \phi_0(x) \hat{O}(x)} \right>=e^{iS_{5D[\phi]}}|_{\phi(x,z=0)=\phi_0(x)}.
\end{equation}
with the field $\phi$. The field $\phi$ corresponds to the classical solution of the 5D action $S_{5D}$. The boundary value $\phi(x,z=0)$ corresponds to the 4D external source $\phi_0(x)$ \cite{4Dexternalsource1maldacena1999large,4Dexternalsource2gubser1998gauge,4Dexternalsource3witten1998anti} within 5D space. By taking the second derivative of the action $S_{5D}$ with respect to the source $\phi_0$, the correlator $\left< \hat{O}(x)\hat{O}(0)\right>$ can be calculated~\cite{S5Dphi0source1miranda2009black}. In this way, it is possible to get the poles of the correlation functions and the masses. However, as mentioned in ref.~\cite{laoshipionmassNoB1-cao2021thermal}, the poles can be easily obtained by solving the 5D equations of motion under certain boundary conditions.  In this way, one can obtain the screening mass by solving the equations of motion with a spatial momentum, while replacing the spatial momentum with frequency for pole mass and thermal width . For details, please refer to ref.~\cite{laoshipionmassNoB1-cao2021thermal}.

Actually, those screening and pole masses are not independent and they are connected by the dispersion relation. Generally, within a thermal medium in an external magnetic field, the dispersion relation (DR) is represented by
\begin{eqnarray} \label{EDRRRRRRRR}
k_0^2&&=\, u_{\perp}^2 \textbf{\textit{k}}_{\perp}^2+u_{\parallel}^2 \textbf{\textit{k}}_{\parallel}^2+m_{\text{pole}}^2,
\end{eqnarray}
where we denote that $u_{\perp} = u_1 = u_2$ and $u_{\parallel} = u_3$ the velocities of transverse and longitudinal directions respectively. The transverse directions are in $x_1,x_2$ directions, and the longitudinal direction is in $x_3$ direction, along which the magnetic field is.  $\textbf{\textit{k}}_{\perp}$ and $\textbf{\textit{k}}_{\parallel}$ are the transverse and longitudinal momenta respectively. Furthermore, $u_i$ represents the pion sound velocity in the $x_i$ direction, respectively. Therefore, \(m_{\text{scr},\perp}\) and $m_{\text{scr},\parallel}$ can be defined in terms of \(m_{\text{pole}}\) and the velocity of sound in that direction, showing as $m_{\text{scr},\perp}=\frac{m_{\text{pole}}}{u_\perp}$ and $m_{\text{scr},\parallel}=\frac{m_{\text{pole}}}{u_\parallel}$.
As discussed in ref.~\cite{yulang1sheng2021pole} and ref.~\cite{pionmassdingxingwang2018meson}, effective thermal masses will lead to \(u_{\perp} < 1\) and \(u_{\parallel} < 1\). Note also that transversal thermal motion is not as strong as longitudinal motion due to the dimensional reduction of magnetic field, giving \(u_{\perp} <  u_{\parallel}\). Finally, we have \(m_{\text{scr},\perp} >  m_{\text{scr},\parallel} > m_{\text{pole}}\). 

After the above preparation,in the following subsections we will derive the EOMs and their asymptotic expansions in the pseudoscalar channel. Then we will extract the masses and present the numerical results of $m_{\text{scr},\perp}$ , $m_{\text{scr},\parallel}$ and \(m_{\text{pole}}\). \(u_{\perp}/u_{\parallel}\) will also be examined as it serves to measure the anisotropy induced by the magnetic field.

\subsection{EOMs and its asymptotic expansion of pseudo-scalar channel} \label{EOMShejiexizhankai}
As the scalar meson field $S(\chi, z)$ and $\pi(\chi,z)$ is decoupled, the scalar meson field $S(\chi, z)$ can be to 0.
The expression of $X$ in eq.~\eqref{expressofX} is reduced to
\begin{equation} \label{Xquan}
    X=\frac{1}{2}\chi I_2 e^{2i \pi^a t^a}.
\end{equation}
Moreover, the interaction between the pion field and the longitudinal component $\varphi^i$ of the axial-vector field $a^i$ occurs exclusively in the pseudo-scalar channel. To simplify our analysis, we will adopt the following decomposition of the gauge field for convenience.
\begin{subequations} \label{zhoushifenliang}
    \begin{equation}
    a_{\mu}^i=a_\mu^{T,i}+\partial_\mu \varphi^i,
\end{equation}
\begin{equation}
    \partial^\mu a_\mu^{T,i}=0.
\end{equation}
\end{subequations}

When obtaining the EOMs within our model, we treat the $\pi^0$ meson as a perturbation and therefore neglect terms beyond quadratic order. Additionally, we adopt the \(A_z = 0\). By combining equations eq.~(\ref{zuoyongliang}), eq.~(\ref{Xquan}) and eq.~(\ref{zhoushifenliang}) and taking the Fourier transformation,
\begin{subequations}
\begin{equation}
      \pi^a(x,z)=\frac{1}{(2\pi)^4}\int dk^4\, e^{-ikx}\pi^a(k,z),
\end{equation}
\begin{equation}
      \varphi^a(x,z)=\frac{1}{(2\pi)^4}\int dk^4\, e^{-ikx}\varphi^a(k,z).
\end{equation}
\end{subequations}
we derive the EOMs governing the behavior of the neutral $\pi^0$ meson and $\varphi$
\begin{subequations} \label{piphieqs}
    \begin{eqnarray}
    \partial_z(\sqrt{g} e^{-\Phi} g^{zz}\chi^2\partial_z\pi)-k_{\mu}^2\sqrt{g}e^{-\Phi}g^{\mu \mu}\chi^2(\pi-\varphi)=0,\nonumber\\
    \\
    \partial_z(\sqrt{g} e^{-\Phi} g^{zz}g^{\mu \mu}\partial_z\varphi)-g_5^2\sqrt{g}e^{-\Phi}g^{\mu \mu}\chi^2(\pi-\varphi)=0,\nonumber\\
\end{eqnarray}
\end{subequations}
where \(k_\mu\) represents the momenta of the neutral pion meson in temporal and different spatial directions with  \(\mu\)= $t, x_1, x_2, x_3$.

From eqs.~(\ref{piphieqs}), since the $x_1$ and $x_2$ directions are perpendicular to the magnetic field, the $x_1$ and $x_2$ (transverse) directions possess isotropic properties. Clearly, for the spatial \(x_1\) and \(x_2\) directions, the system of differential equations are
\begin{subequations} \label{x1piphi}
    \begin{eqnarray}
    \varphi''+&& \left(A'+\frac{f'}{f}+\frac{q'}{2q} -\Phi' \right)\varphi'\nonumber\\
    &&+\frac{e^{2A}g_5^2 \chi^2}{f}(\pi-\varphi)=0,\\    \pi''+&&\left(3A'+\frac{f'}{f}+\frac{q'}{2q}+\frac{h'}{h}-\Phi'+\frac{2\chi'}{\chi} \right)\pi'\nonumber\\
    &&-\frac{k_{1}^2}{fh}(\pi-\varphi)=0.
\end{eqnarray}
\end{subequations}  
Similarly, in the \(x_3\) spatial (longitudinal) direction, the system of differential equations for \(\pi\) and \(\varphi\) are
\begin{subequations} \label{x3piphi}
    \begin{eqnarray}
    \varphi''+&&\left(A'+\frac{f'}{f}-\frac{q'}{2q}+\frac{h'}{h}-\Phi' \right)\varphi'\nonumber \\
    &&+\frac{e^{2A}g_5^2 \chi^2}{f}(\pi-\varphi)=0,\\    \pi''+&&\left(3A'+\frac{f'}{f}+\frac{q'}{2q}+\frac{h'}{h}-\Phi'+\frac{2\chi'}{\chi} \right)\pi'\nonumber\\
    &&-\frac{k_{3}^2}{f q}(\pi-\varphi)=0.
\end{eqnarray}
\end{subequations}  
In the temporal direction, the system of differential equations for \(\pi\) and \(\varphi\) are
\begin{subequations} \label{tpiphi}
    \begin{eqnarray}
    \varphi''+&&\left(A'+\frac{h'}{h}+\frac{q'}{2q}-\Phi'\right)\varphi'\nonumber\\
    &&+\frac{e^{2A}g_5^2\chi^2}{f}(\pi-\varphi)=0, 
    \\
    \pi''+&&\left(3A'+\frac{f'}{f}+\frac{q'}{2q}+\frac{h'}{h}-\Phi'+\frac{2\chi'}{\chi} \right)\pi'\nonumber\\
    &&+\frac{k_{0}^2}{f^2}(\pi-\varphi)=0.
\end{eqnarray}
\end{subequations} 

To solve the eqs.~(\ref{x1piphi}), (\ref{x3piphi}), and(\ref{tpiphi}), we  can also adopt the ``Shooting method''. This numerical approach enables us to tackle these differential equations effectively. In our numerical solution process, we initiate by establishing the asymptotic expansions of the equations at their respective boundaries $z=0$ and $z=z_h$.~\footnote{Due to the metric functions  and $\chi$ exist in eqs~\eqref{x1piphi}-\eqref{tpiphi}, when seeking their asymptotic expansions at the UV and IR boundaries, we must also take into account the asymptotic expansions of eqs.~(\ref{duguiquan}) and~(\ref{chieom}) at the UV and IR regimes, respectively.}. At horizon, one should consider the incoming wave condition~\cite{xuanmingelinhanshuson2002minkowski}. For the spatial direction along $x_1$ and $x_2$, the asymptotic expansion of eqs.~(\ref{x1piphi}) at both boundaries can be described as 
\begin{subequations} \label{x1zhankai}
\begin{eqnarray} \label{x1zhankai1}
        \pi(z\rightarrow 0)=&&\pi_{b0}+\varphi_0+\pi_2 z^2+\frac{\pi_{b0} k_{1}^2}{2h_0}z^2 \ln(z) \nonumber \\ &&+\mathcal{O}(z^3), \\ \label{x1zhankai2}
        \varphi(z\rightarrow 0)=&&\varphi_0+\varphi_2 z^2-\frac{1}{2}g_5^2 m_q^2\zeta^2\pi_{b0}z^2\ln(z)\nonumber \\ && +\mathcal{O}(z^3),\\ \label{x1zhankai3}
        \pi(z\rightarrow z_h)=&&\pi_{h0}+\frac{\pi_{h0}-\varphi_{h0}}{f_{h1}h_{h0}}k_{1}^2(z-z_h)\nonumber \\ &&+\mathcal{O}(z-z_h)^2,\\
        \label{x1zhankai4}
        \varphi(z\rightarrow z_h)=&& \varphi_{h0}+\frac{\varphi_{h0}-\pi_{h0}}{f_{h1}z_h^2}g_5^2\chi_{zh0}^2(z-z_h)\nonumber\\
        &&+\mathcal{O}(z-z_h)^2.
    \end{eqnarray}
\end{subequations}
where $\pi_{b0}$, $\varphi_0$, $\pi_2$, and $\varphi_2$ are the integration constants for the UV boundary, and $\pi_{h0}$ and $\varphi_{h0}$ are the integration constants for the horizon.

For the EOMs in the longitudinal direction ($x_3$ direction), we can obtain their asymptotic expansions of eqs.~(\ref{x3piphi}) at both boundaries as
\begin{subequations} \label{x3zhankai}
\begin{eqnarray}
        \pi(z\rightarrow 0)=&&\pi_{b0}+\varphi_0+\pi_2 z^2+\frac{\pi_{b0} k_{3}^2}{2q_0}z^2 \ln(z) \nonumber \\
        &&+\mathcal{O}(z^3), \\      \varphi(z\rightarrow 0)=&&\varphi_0+\varphi_2 z^2-\frac{1}{2}g_5^2 m_q^2\zeta^2\pi_{b0}z^2\ln(z)\nonumber\\
        &&+\mathcal{O}(z^3),\\        \pi(z\rightarrow z_h)=&&\pi_{h0}+\frac{\pi_{h0}-\varphi_{h0}}{f_{h1}q_{h0}}k_{3}^2(z-z_h)+\nonumber \\
        && +\mathcal{O}(z-z_h)^2,
    \end{eqnarray}
    \begin{eqnarray}
        \varphi(z\rightarrow z_h)=&& \varphi_{h0}+\frac{\varphi_{h0}-\pi_{h0}}{f_{h1}z_h^2}g_5^2\chi_{zh0}^2(z-z_h)\nonumber\\
        &&+\mathcal{O}(z-z_h)^2.
    \end{eqnarray}
\end{subequations}
where $\pi_{b0}$, $\varphi_0$, $\pi_2$, and $\varphi_2$ are the integration constants for the UV boundary, and $\pi_{h0}$ and $\varphi_{h0}$ are the integration constants for the horizon. 

Similarly, in the temporal direction, we can obtain the asymptotic expansion of eqs.~(\ref{tpiphi}) at both boundaries, given by
\begin{subequations} \label{tzhankai}
\begin{eqnarray}
        \pi(z\rightarrow 0)=&&\pi_{b0}+\varphi_0+\pi_2 z^2-\frac{\pi_{b0} k_{0}^2}{2}z^2 \ln(z) \nonumber \\
        && +\mathcal{O}(z^3), \\
        \varphi(z\rightarrow 0)=&&\varphi_0+\varphi_2 z^2-\frac{1}{2}g_5^2 m_q^2\zeta^2\pi_{b0}z^2\ln(z)\nonumber \\
        && +\mathcal{O}(z^5),\\ \label{tzhankaipimodel2}
        \pi(z\rightarrow z_h)=&&\varphi_{h0}+(z-z_h)^{\frac{i k_0}{f_{h1}}} \Big[\pi_{h0}\nonumber\\
        &&+\mathcal{O}(z-z_h)^1 \Big],\\
        \varphi(z\rightarrow z_h)=&& \varphi_{h0}+(z-z_h)^{\frac{i k_0}{f_{h1}}}\nonumber\\
        &&\times\Big[ \frac{i g_5^2 \pi _{h0} f_{{h1}} \chi _{{zh0}}^2}{{z_h}^2 k _t \left(f_{{h1}}+i k _t\right)}(z-z_h)\nonumber\\
        && +\mathcal{O}(z-z_h)^2 \Big].
    \end{eqnarray}
\end{subequations}
The $\pi_{b0}$, $\varphi_0$, $\pi_2$, and $\varphi_2$ are the integration constants for the UV boundary, and $\varphi_{h0}$ and $\pi_{h0}$ are the integration constants for horizon. 

As explained in ref.~\cite{xuanminprdjiezicao2020pion}, due to the presence of the term $(\pi-\varphi)$ in eqs.~(\ref{piphieqs}), if one makes a transformation of assuming $\pi_c=\pi+c$ and $\varphi_c=\varphi+c$ with $c$ a nonzero constant, one can find that the solution is still available.  Consequently, we can take integration constant \(\varphi_0\) as an extraneous free parameter. For convenience, we let $\varphi_0=0$. Additionally, due to the linear nature of eqs.~(\ref{piphieqs}), for the integral constants \(\pi_{h0}\) at the IR boundary, we can set it to unity. 

With the asymptotic expansion solutions, we can use the "shooting method" to numerically solve  the differential equation system~(\ref{piphieqs}) for \(\pi\) and \(\varphi\). Since the second order differential equations, one should matching the function values and its first order derivatives. Therefore, there are four constraints. However, as per the analysis presented above, the integration constants that we need to solve are five: $\pi_{b0}$ \(\pi_2\), \(\varphi_2\), \(\varphi_{h0}\) and the \(k_i\) for each direction. However, a physically analyzing of the solutions of differential equations and two-point Retarded correlator allow us to determine another integration constant, \(\pi_{b0}\), for each direction, which will be explored and discussed in the subsequent subsections.

\begin{figure*}[htb]
\centering
    \begin{overpic}[scale=0.61]{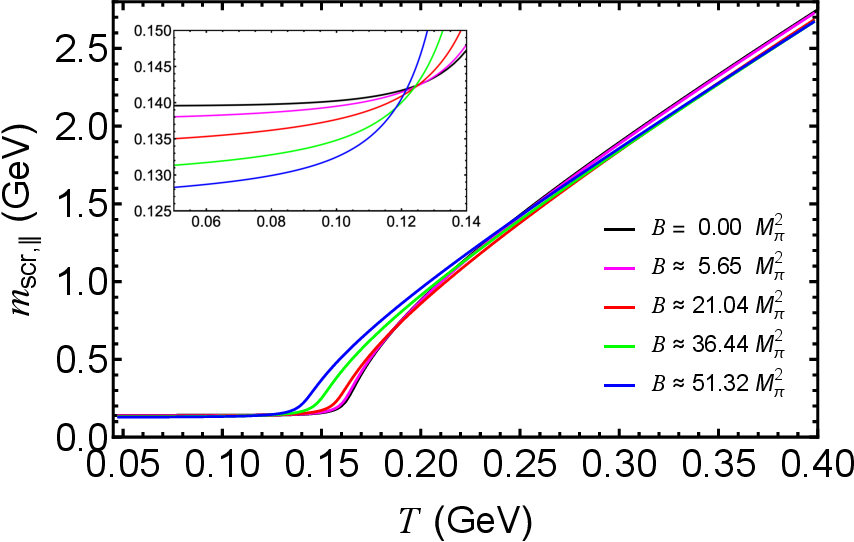}
        \put(80,58){\bf{(a)}}
    \end{overpic}
    \begin{overpic}[scale=0.61]{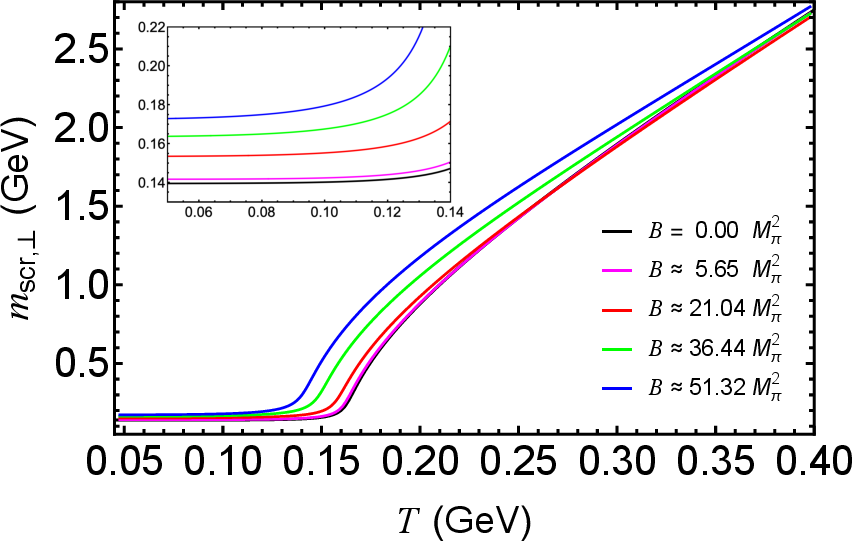}
        \put(80,58){\bf{(b)}}
    \end{overpic}
\caption{The plots illustrate the \(m_{\text{scr},\parallel}\) and $m_{\text{scr},\perp}$ as a function of \(T\) under different fixed \(B\) respectively. 
The fixed magnetic field \(B\) is about 0, 5.56, 21.04, 36.44 and 51.32 \(M_{\pi}^2\), which corresponds to 0, 0.11, 0.41, 0.71, and 1 GeV\(^2\), respectively in our model. 
The inset in (a) and (b) is a zoom for the range $T \in [0.05, 0.14 ]$ GeV. For the $B=$ 0, 5.56, 21.04, 36.44, 51.32 $M_{\pi}^2$, their corresponding $T_{\text{pc}}$ of chiral phase transition are 161.0, 160.2, 156.0, 146.5, 138.8 MeV, respectively.} 
\label{twomodelsx1x2x3screenwithTT}
\end{figure*}

\begin{figure*}[htb]
\centering
    \begin{overpic}[scale=0.61]{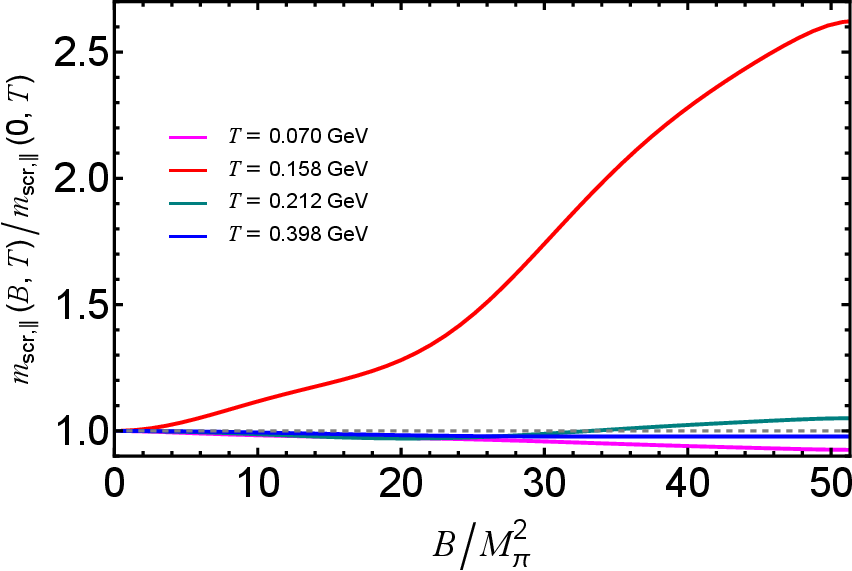}
        \put(59,63){\bf{(a)}}
    \end{overpic}
    \begin{overpic}[scale=0.61]{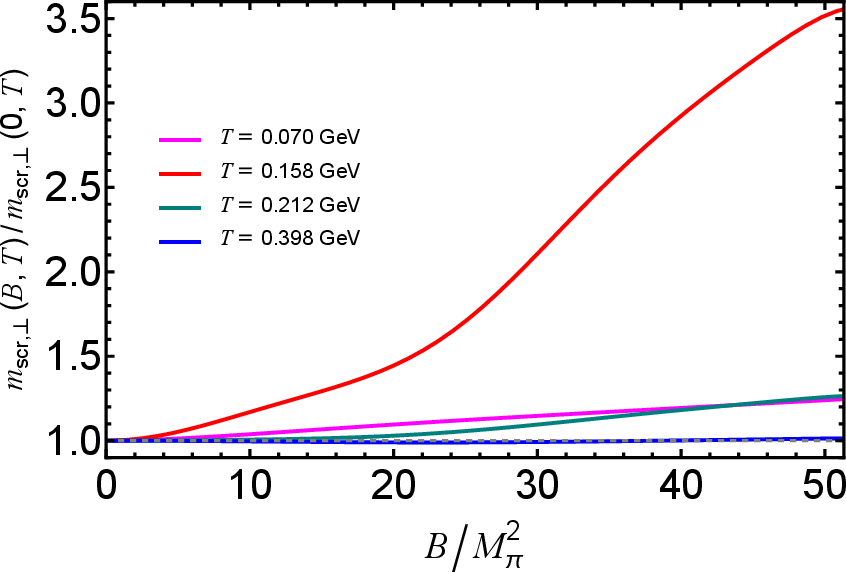}
        \put(60,63){\bf{(b)}}
    \end{overpic}
\caption{The plots depict normalized screening mass, $m_{\text{scr}}(B,T)/m_{\text{scr}}(0,T)$, as a function of \(B/M_{\pi}^2\) under different fixed \(T\), where the temperatures \(T\) are chosen as 70, 158, 212 and 398 MeV, respectively. The figures (a) and (b) are described for normalized screening masses in longitudinal and transverse screening masses, respectively.}
\label{twomodelsx1x2x3NormalizedscreenwithBB}
\end{figure*}

\subsection{Numerical results of screening masses  of neutral pion}

In this subsection we will proceed to numerically solve the EOMs for \(\pi\) and \(\varphi\), eqs.~\eqref{x1piphi} and ~\eqref{x3piphi}. Then we will show the transverse and longitudinal screening masses of the neutral pion.

Following the holographic dictionary, $\pi_{b0}$, as shown in the asymptotic expansions in eqs.~\eqref{x1zhankai} and~\eqref{x3zhankai}, is interpreted as the external sources $J_{\pi}$. Actually, considering the spatial correlator in eq.~(\ref{screenyinyong}), the desired values of \(m_{\text{scr},\perp}\) or $m_{\text{scr},\parallel}$ for the neutral pion meson is situated at the pole of the Green's function, representing its singularities. Consequently, at the UV boundary, the integral constants $\pi_{b0}$ in the asymptotic expansion can be set to zero~\cite{xuanminprdjiezicao2020pion,Liang:2023lgs}. Thus, the $m_{\text{scr},\perp}$ and $m_{\text{scr},\parallel}$ are the imaginary part of $k_{1}$ and $k_{3}$.

After numerically solving the EOMs of \(\pi\) and \(\varphi\) in different directions by combining eqs.(\ref{duguiquan}) and~(\ref{chieom}), we show the solution of the screening mass of the neutral pion in Figs.~\ref{twomodelsx1x2x3screenwithTT} and~\ref{twomodelsx1x2x3NormalizedscreenwithBB}.

The behavior of the screening masses with respect to $T$ at different fixed $B$ is shown in Fig.~\ref{twomodelsx1x2x3screenwithTT}. At low temperature, \(m_{\text{scr},\parallel}\) and \(m_{\text{scr},\perp}\) are both slightly affected by $T$. As shown in the inset, $m_{\text{scr},\parallel}$ and $m_{\text{scr},\perp}$ show decreasing and increasing behaviors with $B$ at low temperature region. In the critical region, close to $T_{\text{pc}}(B)$, both screening masses suddenly increase.  As the temperature increases, the two screen masses are characterized by the linear thermal mass and tend to be the same for different given $B$ at extremely high temperatures.

\begin{figure*}[htb]
\centering
    \begin{overpic}[scale=0.615]{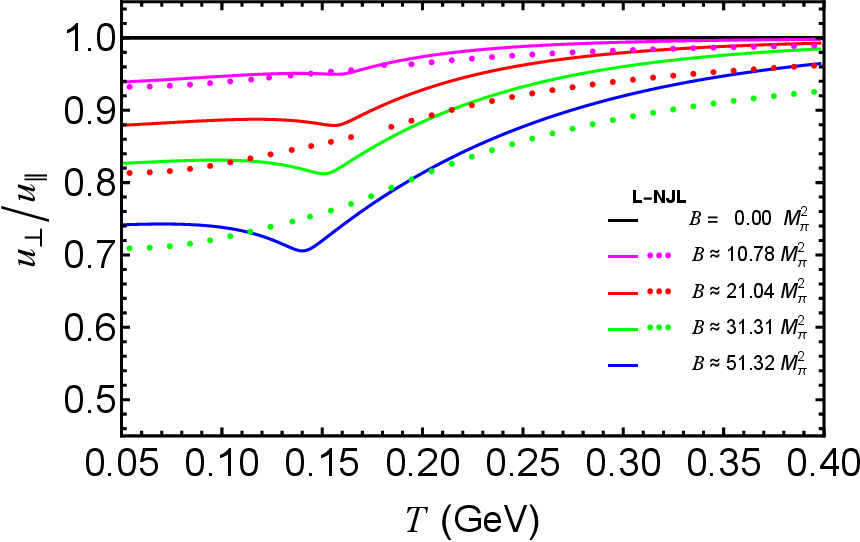}
        \put(50,20){\bf{(a)}}
    \end{overpic}
    \begin{overpic}[scale=0.604]{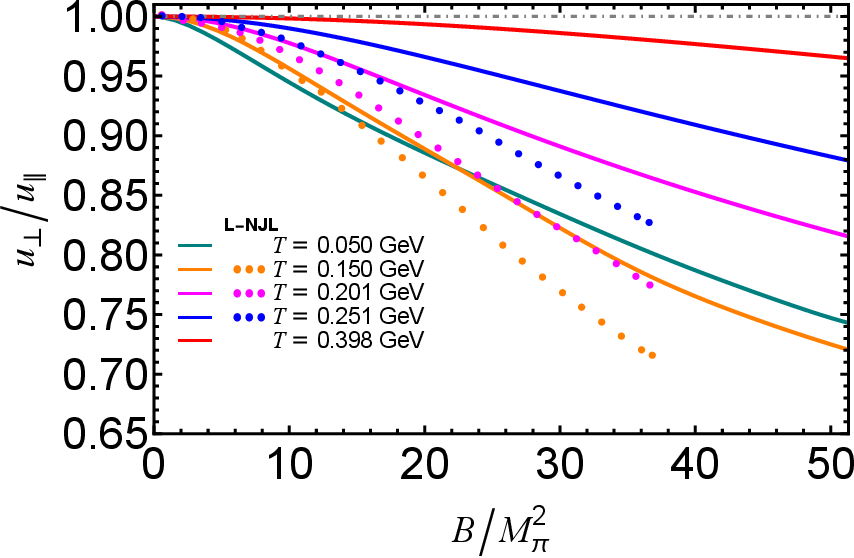}
        \put(60,20){\bf{(b)}}
    \end{overpic}
\caption{(a), the plot illustrates the ratio of sound velocity $u_{\perp}/u_{\parallel}$ as a function of \(T\) under different fixed \(B\). The range of \(B\) is approximately from 0 to 51.32 \(M_{\pi}^2\), which corresponds to 0 to 1 GeV\(^2\) in our model. (b), it depicts the ratio of sound velocity $u_{\perp}/u_{\parallel}$ as a function of \(B/M_{\pi}^2\) under different fixed \(T\), where the temperatures \(T\) were chosen as 50, 150, 201, 251 and 398 MeV, respectively. For the $B=$ 0, 10.78, 21.04, 31.31, 51.32 $M_{\pi}^2$, their corresponding $T_{\text{pc}}$ of chiral phase transition are 161.0, 158.7, 156.1, 150.2, 138.8 MeV respectively. Furthermore, we have extracted results from the lattice-improved NJL model~\cite{yulanglatticeNJLsheng2022impacts}, represented by dots in the graph. The displayed results from lattice-improved NJL model have been scaled according to $M_{\pi}$ of its model \(M_{\pi} = 138\) MeV.}
\label{twomodelssoundspeedwithTTTandBBB}
\end{figure*}

In Fig.~\ref{twomodelsx1x2x3NormalizedscreenwithBB}, we plot the normalized screening mass of neutral pion mesons with $B$ at different fixed $T$. As shown, the normalized screening masses ($m_{\text{scr},\parallel}$ and $m_{\text{scr},\perp}$), either at low ($T=70$) MeV or at higher temperatures ($T$= 212 and 398 MeV), depend slightly on $B$, which corresponds to the main properties of a neutral particle. More precisely, at $T=70$ MeV the $m_{\text{scr},\parallel}$ decrease with increasing $B$, but the $m_{\text{scr},\perp}$ increase with increasing $B$. At $T=212$ MeV, $m_{\text{scr},\parallel}$ first decrease and then start to increase as $B$ grows. And they always increase with $B$ when the temperature is $T<398$ MeV. At a higher given temperature, $T=398$ MeV,  $m_{\text{scr},\parallel}$ show the behavior of decreasing with $B$ again, and the decrease is very small. This is in agreement with the lattice result~\cite{latticepionding2022chiral}. However, at a certain temperature where $T=158$ MeV near $T_{\text{pc}}(B)$, the normalized screening masses increase significantly with $B$, which can be understood as the magnetic dependence being enhanced by the critical fluctuation.

From the energy dispersion relation in eq.~\eqref{EDRRRRRRRR}, It is known that the ratio \(u_{\perp}/u_{\parallel}=m_{\text{scr},\parallel}/m_{\text{scr},\perp}\). In Fig.~\ref{twomodelssoundspeedwithTTTandBBB}(a), we present the $T$ dependence of the ratio $u_{\perp}/u_{\parallel}$ at different $eB$. For finite $eB$, $u_{\perp}/u_{\parallel}$ is almost independent of temperature at low $T$. As the temperature increases to near $T_{\text{pc}}(B)$, $u_{\perp}/u_{\parallel}$ show a non-monotonic bump, and the bump is more obvious as $eB$ increases. When the temperature is $T>T_{\text{pc}}(B)$, all curves increase with increasing $T$. When $T \gg T_{\text{pc}}(B)$, all curves approach 1. Mathematically, increasing $T$ will cause both $q(z)$ and $h(z)$ to approach 1, as shown in the subsection~\ref{duguiquanjietu}, and then it induces that eq.~(\ref{x1piphi}) is almost the same as eq.~(\ref{x3piphi}). This indicates that in all three spatial dimensions, the thermal fluctuations dominate over the anisotropy induced by the magnetic field. Furthermore, in Fig.~\ref{twomodelssoundspeedwithTTTandBBB}(b), we study the $B$ dependence of the ratio \(u_{\perp}/u_{\parallel}\). The ratio \(u_{\perp}/u_{\parallel}\) decreases with increasing $B$, which is the consequence of the decoupling of the transverse dimension in the strong limit of the magnetic field. Again, in the region of the critical temperature, the curve decays faster with $B$, as expected, as shown by the orange line. 

\subsection{Numerical results of pole mass and thermal widths of neutral pion}

\begin{figure*}[htb]
\centering
    \begin{overpic}[scale=0.615]{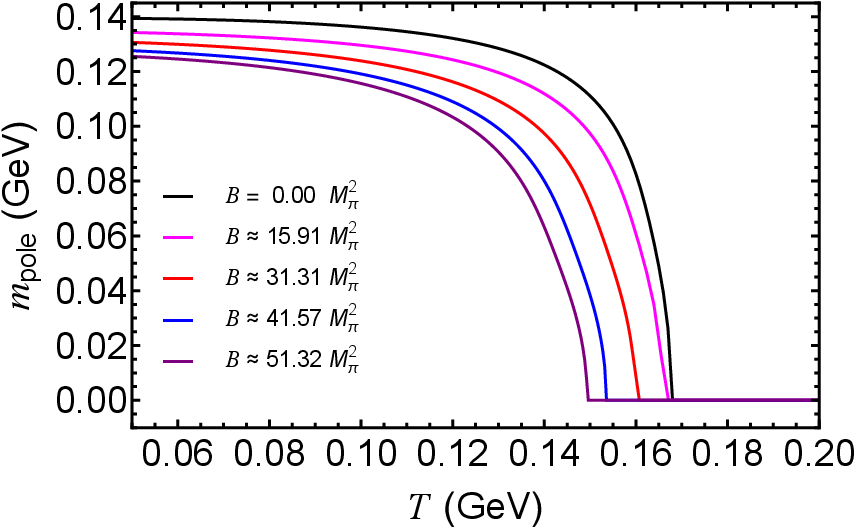}
        \put(85,55){\bf{(a)}}
    \end{overpic}
    \begin{overpic}[scale=0.61]{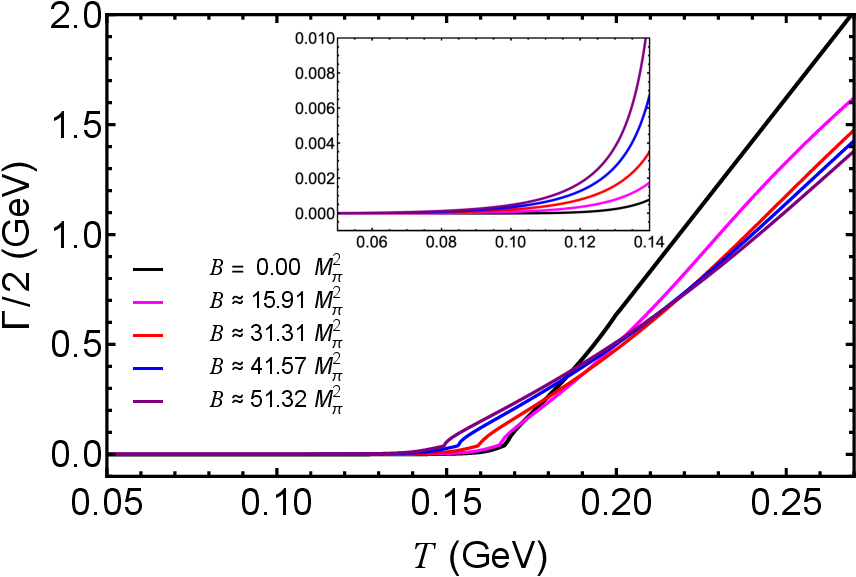}
        \put(82,55){\bf{(b)}}
    \end{overpic}
\caption{The plots illustrate the \(m_{\text{pole}}\) and $\Gamma/2$ as a function of \(T\) under different fixed \(B\) respectively. The selection of \(B\) is about 0, 15.91, 31.31, 41.57 and 51.32 \(M_{\pi}^2\), which corresponds to 0, 0.51, 0.81 and 1 GeV\(^2\) respectively in our model. The inset in (b) is a zoom for the range $T \in [0.05, 0.14 ]$ GeV.}
\label{twomodelspoleandrekuanduwithTTT}
\end{figure*}

\begin{figure*}[htb]
\centering
    \begin{overpic}[scale=0.625]{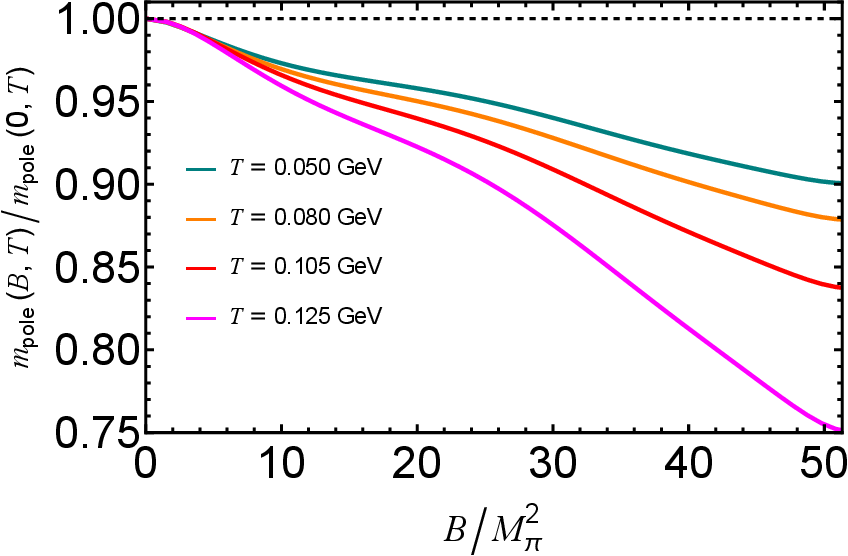}
        \put(20,55){\bf{(a)}}
    \end{overpic}
    \begin{overpic}[scale=0.605]{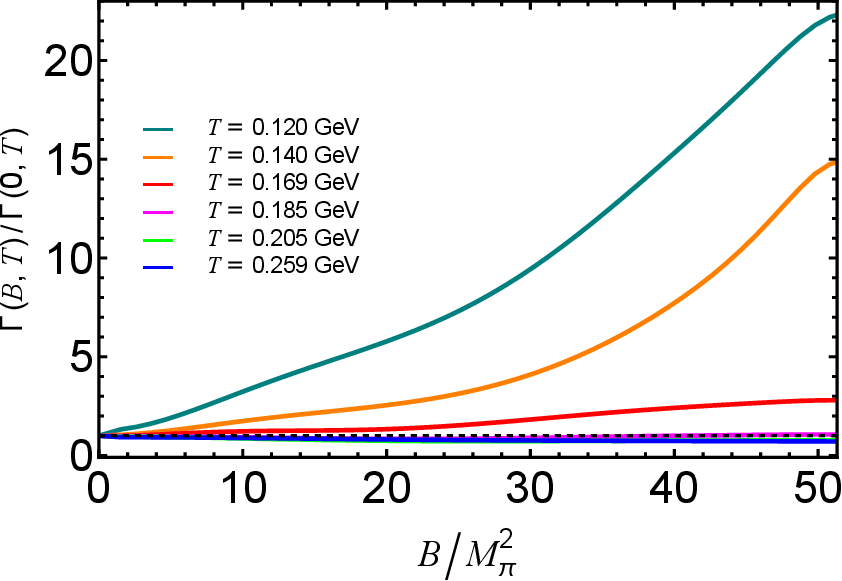}
        \put(15,60){\bf{(b)}}
    \end{overpic}
\caption{The plots depict normalized \(m_{\text{pole}}\) and normalized \(\Gamma/2\) as a function of \(B/M_{\pi}^2\) under different fixed \(T\), where the temperatures \(T\) are chosen as 50, 80, 105, 125 MeV, and 120, 140, 169, 185, 205, 299 MeV respectively.}
\label{twomodelsNormalizedpoleandrekuanduwithBBB}
\end{figure*}

In this subsection, we present the behavior of the pole mass~$m_{\text{pole}}$ and thermal width~$\Gamma/2$ of the neutral pion meson at given $T$ and $B$. 

For asymptotic expansion in eq.~\eqref{tzhankai}, according to the holographic dictionary, the integration constant \(\pi_{b0}\) is understood to correspond to the external sources denoted as \(J_{\pi}\). 
From the temporal correlator eq.~(\ref{poleyinyong}), we know that the complex frequencies \(k_0\) precisely lie in the location of the poles in the Green's function. Evidently, this corresponds exactly to the position of singularities. Thus, for the integration constant \(\pi_{b0}\), its value must be 0~\cite{xuanminprdjiezicao2020pion,Liang:2023lgs}. Hence, taking into account the analysis from the Section~(\ref{EOMShejiexizhankai}), we have already determined that \(\varphi_0=0\), \(\pi_{b0}=0\), and \(\pi_{h0}=1\). Similarly, we employ the "shooting method" to numerically solve the differential eqs.~(\ref{tpiphi}). Thus, the $m_{\text{pole}}$ and thermal width are the real and imaginary part of $k_0$, respectively.

In Fig.~\ref{twomodelspoleandrekuanduwithTTT}, we study the temperature dependence of \(m_{\text{pole}}\) and thermal widths \(\Gamma/2\) of the neutral pion meson under different fixed $B$. From Fig.~\ref{twomodelspoleandrekuanduwithTTT}(a), \(m_{\text{pole}}\) decrease with temperature and go to zero near \(T_{\text{pc}}\). On the contrary, in Fig.~\ref{twomodelspoleandrekuanduwithTTT}(b), for given $eB$, the thermal widths (\(\Gamma/2\)) are small at low temperatures. And then, in the region above the critical temperature, they increase rapidly with the increase of $T$. Since the thermal width is related to the dissociation level, such curves naively indicate that $\pi^0$ is a well-behaved bound state at low $T$ and becomes loose at high $T$. Indeed, without $B$, the corresponding behavior of the real and imaginary parts of the neutral pion propagator is consistent with results from LQCD simulations~\cite{polexiajiangbrandt2015pion}, $\chi$PT~\cite{poxiajiangCPTson2002real} as well as holography results~\cite{laoshipionmassNoB1-cao2021thermal,xuanminprdjiezicao2020pion}.

In Fig.~\ref{twomodelsNormalizedpoleandrekuanduwithBBB}, we have extracted the normalized \(m_{\text{pole}}\) and \(\Gamma/2\) as functions of $B/M_{\pi}^2$ at different fixed $T$. Clearly, in Fig.~(\ref{twomodelsNormalizedpoleandrekuanduwithBBB}a), the normalized \(m_{\text{pole}}\) decreases with increasing $B$ at finite temperatures. In vacuum, such a similar behavior has already been found in lattice simulations~\cite{piondianlizinoding2021chiral} and studied by model calculations~\cite{Xing:2021kbw,Lin:2022ied,Chao:2022bbv}. Meanwhile, in Fig.~\ref{twomodelsNormalizedpoleandrekuanduwithBBB}(b), we observe that when $T\lesssim T_{\text{pc}}(B)$, \(\Gamma/2\) exhibits strong dependence on the magnetic field and increases with $B$. When $T \gtrsim T_{\text{pc}}(B)$, \(\Gamma/2\) is slightly affected by the magnetic field and the thermal width is almost controlled by the magnitude of the temperatures. We conclude that the magnetic field is more involved in the thermal effects below than above the critical temperature, i.e. temperature and magnetic field are mostly entangled at moderate $T$. Moreover, at high temperature, $T \gg T_{\text{pc}}(B)$, \(\Gamma/2\) decrease with increasing $B$. It is interesting to note that, at high temperature, the behavior of \(\Gamma/2\) is similar with $m_{\text{scr},\parallel}$.

\section{Conclusion and discussion} \label{summaryanddiscussion}

In this work,  we mainly investigate the thermal properties of neutral pions within a hot and magnetized medium, including their screening masses, pole masses, and thermal widths, by using a lattice-improved soft-wall AdS/QCD model. 

We introduce the magnetic field within the Einstein-Maxwell holographic model following refs.~\cite{yinruB1mamo2015inverse,yinruB2dudal2016no,laoshishouzhengB1=li2017inverse}. By numerically solving the equations of motion, we obtain the full solutions with constant magnetic field. Compared with the expanding solutions in refs.~\cite{yinruB1mamo2015inverse,yinruB2dudal2016no,laoshishouzhengB1=li2017inverse}, the full solutions turn out to be valid in a wider range of temperature and magnetic field. Based on those numerical solutions, we consider the chiral dynamics in the soft-wall AdS/QCD model. By considering an effective coupling between the dilaton field and the flavor sector and introducing a magnetic-field-dependent coupling function \(g(B)\), which is fitted by results for \(T_{\text{pc}}(B)\) from lattice simulation ~\cite{chiralTcpIMClatticebali2012qcd}, we successfully capture the MC effect of chiral condensation at low temperatures and the IMC effect near the chiral transition temperature \(T_{\text{pc}}(B)\).  It is worth of mentioning that a better fitting of the chiral transition temperature \(T_{\text{pc}}(B)\) has been obtained, and all the other qualitative behaviors are in good agreement with the lattice simulations results~\cite{chiralMCandIMC1latticebali2012qcd},   which shows a good description of chiral phase transition from such an effective holographic QCD model.

Then we study the thermal properties of the neutral pion, which is still the Goldstone boson of the chiral symmetry breaking even within a magnetic field. The temperature and magnetic field dependence of the screening masses $m_{\text{scr},\parallel}$ and $m_{\text{scr},\perp}$, characterizing the mesonic correlations in perpendicular and vertical directions respectively, are extracted. If one fixes the magnetic field, both $m_{\text{scr},\parallel}$ and $m_{\text{scr},\perp}$ are slightly affected by $T$ at low temperature. In the temperature region close to transition temperatures $T_{\text{pc}}(B)$, the screening masses significantly increase. Such enhanced behavior shows the connection between the screening masses and the chiral phase transition. After the phase transition, the mesons are not tightly bounded and the hot and magnetized medium start imposing strongly impact on their properties. Then, when the temperature increases further, the two screening masses are shown to increase linearly with $T$ and tend to be the same for different given $B$ at extremely high temperatures. Those results are consistent with the results from effective models like NJL, confirming the effectiveness of the soft-wall holographic description on chiral dynamics.

Furthermore, we fixed temperature and study the $B$ dependence of screening masses. Roughly speaking, the screening masses do not change much below $T_{\text{pc}}$. This is reasonable, since the neutral pions from our holographic model are still tightly bound states and as neutral particles their properties would not be changed significantly by the magnetic field. But if we look carefully into the results, we can find that at low temperature $m_{\text{scr},\parallel}$ decreases as $B$ grows, while $m_{\text{scr},\perp}$ increases, showing an anisotropic effect of the magnetic field at low temperature. Such a qualitative behavior can also be found from NJL model calculation \cite{yulanglatticeNJLsheng2022impacts}. Then, when the temperature is taken to be near $T_{\text{pc}}$, both $m_{\text{scr},\parallel}$ and $m_{\text{scr},\perp}$ increase with $B$. At high temperature region, $m_{\text{scr},\perp}$ still increase as $B$ grow but $m_{\text{scr},\parallel}$ show behavior of decreasing with increasing $B$ again. It is worth mentioning that a similar behavior can be seen in the lattice study \cite{latticepionding2022chiral}. Moreover, we also study the ratio \(u_{\perp}/u_{\parallel}\). When $T \gg T_{\text{pc}}(B)$, all the curves approach to 1. This indicates that the difference between \(u_{\perp}\) and \(u_{\parallel}\) is vanishing. For the $B$ dependence of the ratio \(u_{\perp}/u_{\parallel}\). The ratio \(u_{\perp}/u_{\parallel}\) decrease with the increase of $B$.  This indicates the destruction of spatial symmetry by the magnetic field. We would like to emphasize that by simply fitting data of $T_{\text{pc}}(B)$ from lattice simulation, all the qualitative behaviors for the pion properties can be consistent with lattice simulation simultaneously. In some sense, it confirms the effectiveness of holographic QCD method.

The screening mass depicts the spatial effect, and we have seen that it can be well described by the holographic method. As for the temporal effect on the pionic sector, there are still not many data from lattice simulations. But the extension in hologarphic method is quite directly. We can obtain the poles in complex frequency plane and get the pole mass \(m_{\text{pole}}\) and the thermal widths \(\Gamma/2\) from their real and imaginary part.  If we fix $B$, we find that \(m_{\text{pole}}\) decreases with temperature and goes to zero near \(T_{\text{pc}}\), while the thermal widths \(\Gamma/2\) increase monotonically with increasing $T$. It is interesting to see the pole mass decrease below $T_\text{pc}$, which are consistent with the analysis from scaling law in the finite temperature chiral perturbation theory in \cite{son2002pion}. Also, it is consistent with the lattice study in \cite{polexiajiangbrandt2015pion}, in which the mass of pions are taken slightly higher than their physical values. Finally, at different fixed temperatures,  $m_{\text{pole}}$ monotonically decrease with $B$ upto $1\rm{GeV}^2$. This behavior is qualitatively consistent with the lattice-improved NJL study in \cite{yulanglatticeNJLsheng2022impacts} while it differs from the NJL calculation in ref.~\cite{Li:2020hlp}, in which $m_{\text{pole}}$ decreases when $B<0.8\rm{GeV}^2$ and increases when $B>0.8 \rm{GeV}^2$. Also, being similar with the behavior of $m_{\text{scr},\parallel}$,  we can see that the thermal widths of the $\pi^0$ slightly decrease as $B$ grows above the critical temperature. In this work, we found that the thermal properties of magnetized pions are mainly determined by the temperature itself at high $T$. These are preliminary results of the dependence on the magnetic field, since such a weak dependence would be easily washed out by other effects, for example the backaction from the metric, and a full calculation will be required in further work.

In the present study, we see that by fitting our holographic model using the condensation data from lattice, the model can capture most of the qualitative behaviors, especially for the thermal properties of pions. Due to the limitation from our current numerical techniques, we can not compare the zero temperature behavior of the pole masses under magnetic field and check the main difference between our results and the NJL studies ref.~\cite{Li:2020hlp}. We will leave it to the future. Furthermore, the recent LQCD simulation results indicate that within the $B$ range of approximately 4 to 9 \(\text{GeV}^2\), chiral condensation switches from a crossover to first order at a critical endpoint located in this range~\cite{d2022phase}. It is also interesting to extend our study to stronger magnetic field case.

\section{Acknowledgments}
NW thanks Wei-Jiang Liang for valuable discussions. This work is supported by the National Natural Science Foundation of China (NSFC) under Grant Nos. 12275108, 12305142, Science and Technology Planning Project of Guangzhou, China under Grant No. 2024A04J3243 and the Fundamental Research Funds for the Central Universities under Grant No. 21622324. JC is supported by the National Natural Science Foundation of China (NSFC) under Grant No. 12365020.

\bibliography{refs.bib}

\end{document}